\documentclass[final]{jfm}

\usepackage{graphicx}
\usepackage{newtxtext}
\usepackage{newtxmath}
\usepackage{natbib}
\usepackage{hyperref}
\usepackage{empheq}
\hypersetup{
    colorlinks = true,
    urlcolor   = blue,
    citecolor  = black,
}

\newcommand{\RomanNumeralCaps}[1]
\linenumbers

\newcommand{\mbf}[1]{\mbox{\boldmath{$#1$}}}
\newcommand{\bt}[1]{{\mathbf #1}}

\newcommand{\heli}{{h}}
\newcommand{\ellfive}{{\ell}}
\definecolor{darkgreen}{rgb}{0,0.35,0}

\title{{$(3+1)$-dimensional compressible fluid as a $(4+1)$-dimensional Chern-Simons system}}

\author{Miguel D. Bustamante,\aff{1}\corresp{ 
\email{miguel.bustamante@ucd.ie, laura.andrianopoli@polito.it, mario.trigiante@polito.it, z@cecs.cl, jorge.zanelli@uss.cl}}  Laura Andrianopoli,\aff{2,}\aff{3} Mario Trigiante,\aff{2,}\aff{3} \and Jorge Zanelli\aff{4,}\aff{5} 
}
\affiliation{
\aff{1}School of Mathematics and Statistics, University College Dublin, Belfield, Dublin 4, Ireland
\aff{2}Politecnico di Torino, Corso Duca degli Abruzzi, 24, 10129 Torino, Italy
\aff{3}INFN, Sezione di Torino via P. Giuria 1, 10125 Torino, Italy
\aff{4}Centro de Estudios Cient\'ificos, CECs, Arturo Prat 514, Valdivia, Chile
\aff{5} Universidad San Sebasti\'an, General Lagos 1163, Valdivia, Chile}
\begin{document}
\maketitle

\begin{abstract}
{A fluid described by an Abelian} Chern-Simons action principle in $4+1$ dimensions is considered. Letting $3+1$ dimensions correspond to the usual space and time, and assuming the fields to be independent on the fifth coordinate, the free theory provides an interpretation as a system of advection equations, where the advecting velocity field is defined as the null vector of the field strength tensor (curvature). The free theory possesses a number of conservation laws which turn out to be prototypical forms of helicity and entropy conservation. {Coupling} the Chern-Simons field to an external source, a new conserved charge density is obtained which has the form of the Rossby-Ertel's potential vorticity. Finally, by identifying the external current with the Chern-Simons field {
in a gauge-invariant setting}, based on non-relativistic ideas, a self-interacting action principle is obtained whose Euler-Lagrange equations correspond precisely to a classical dissipationless compressible $(3+1)$-dimensional fluid endowed with thermodynamics, with only one extra condition: a constraint on the initial profile of the Rossby-Ertel's potential vorticity. After analysing this constraint of the ``Chern-Simons fluid formulation'', 
 we investigate the helicity conservation of general fluids, going beyond classical analyses of barotropic fluids and no-cross boundary conditions for vorticity \citep{moffatt1969degree}. A new fluid helicity invariant for barotropic fluids under {generic} boundary conditions is obtained and the role of baroclinity in the helicity production is clarified. 
{Inside a region bounded by an isentropic surface}, the theory's constraint on the Rossby-Ertel potential vorticity gives {an integral} formula for the mass, and for the evolution of fluid helicity in the baroclinic case. Finally, for an ideal gas exact, steady solutions of the equations of motion are found in a rotating scenario, showing that Ferrel-cell like patterns are produced in a rotating planet.   
\end{abstract}

%\begin{keywords}
%Authors should not enter keywords on the manuscript, as these must be chosen by the author during the online submission process and will then be added during the typesetting process (see \href{https://www.cambridge.org/core/journals/journal-of-fluid-mechanics/information/list-of-keywords}{Keyword PDF} for the full list).  Other classifications will be added at the same time.
%\end{keywords}

%{\bf MSC Codes }  {\it(Optional)} Please enter your MSC Codes here

\newpage

\section{Introduction}
\label{sec:headings}
The Chern-Simons action is a functional for a gauge connection, a Lie algebra-valued 1-form $A_\mu(x) \mathrm{d}x^\mu$,  defined on an odd-dimensional manifold $\mathcal{M}_{2n+1}$. This functional does not require a metric and is therefore independent of the signature of the spacetime where the evolution takes place, making it suitable to describe both relativistic and non-relativistic situations. The use of exterior forms renders the Chern-Simons formulation coordinate-independent, naturally generally covariant and equally appropriate for flat or curved spacetimes.

Chern-Simons equations offer one of the most basic forms to represent a dynamically evolving system. Hamilton's equations describing the evolution of a mechanical system in phase space is a particular case, where the corresponding action is a Chern-Simons form for an Abelian connection in $0+1$ dimensions (i.e., the dynamical variables depend on time only, not on space); the coupling of a point charge to the electromagnetic field is another unexpected $0+1$ Chern-Simons system \citep{Zanelli:2008sn}. 

Chern-Simons dynamics is observed in a large class of systems, ranging from those mentioned above to $(2+1)$-dimensional gravity and supergravity \citep{deser1984three,achucarro1986chern}, to high $T_c$ superconductivity \citep{Randjbar-Daemi:1989ett,Wang:2022}, to the shallow-water equations that can model the dynamics of the earth's atmosphere \citep{jacques2022simplified,Tong:2022gpg}. The  wide range of physical applications reflects the robustness, simplicity and versatility of Chern-Simons theories. A notable application in $3$ spatial dimensions is the spatial Chern-Simons $3$-form (also known as the Hopf invariant), which plays a crucial role in the topological interpretation of fluid and magnetic helicity, and in topological fluid mechanics in general \citep{woltjer1958theorem,moffatt1969degree,jackiw2004perfect, arnold2009topological, liu2010kauffman, holm2011geometric, liu2012jones, liu2015derivation, lingam2016concomitant, webb2018helicity}.

Applications of Chern-Simons models in higher odd dimensions have remained as a somewhat academic problem mostly in relation to gravitation, supergravity \citep{cremmer1978supergravity,hassaine2016chern} and string theory \citep{dimofte2013chern}. In this paper we investigate the Chern-Simons paradigm in a ($4+1$)-dimensional spacetime and show that it produces a dynamical system that describes a realistic fluid in $3+1$ dimensions.

It is important to remark that the Chern-Simons action principle presented here stands out when contrasted with the vast classical literature on  variational principles for fluid mechanics, known as the Clebsch formulation. That classical work, was excellently summarised in the review by \cite{salmon1988hamiltonian}, followed by the reviews by \cite{zakharov1997hamiltonian} and \cite{morrison1998hamiltonian}, with plenty of modern applications, such as \citep{rumpf2022generalized}. The main idea of these classical works is that, starting from the variational principle for a discrete system of particles one obtains, in the continuum limit, a variational principle for the Eulerian fields and the equations of motion for a compressible dissipationless fluid. This formulation possesses a natural invariance under the re-labelling of particle positions, which directly leads to conservation laws such as the Rossby-Ertel's potential vorticity. However, these classical formulations have a conceptual disadvantage: for each velocity field component, two scalar variables are required in the Eulerian formulation. Such doubling of degrees of freedom makes the formulation computationally less attractive due to the need for extra  memory and instabilities in the time evolution of the gradients of the new scalar degrees of freedom. Analytically, there is a drawback in the non-uniqueness of the mapping from velocity field to the six scalars due to the fact that the velocity field components are obtained as a nonlinear (quadratic) form in terms of the six scalars of the formulation. Although some attempts have been made to study these equations \citep{ohkitani2003numerical, cartes2007generalized}, one cannot escape the need to manually `reset' these six scalar degrees of freedom every so often in order to cure the instability of the gradients.

In contrast, the Chern-Simons action principle proposed here does not introduce extra degrees of freedom, except for a ``dressing'' scalar gauge field, which is needed so that the theory remains gauge invariant, as is natural for a Chern-Simons theory. This dressing field plays an important role in generalising the fluid helicity to arbitrary boundary conditions. Moreover, the relation between the velocity field and the vector potential of the Chern-Simons theory is linear (affine), so that the mapping from velocity field to vector potential is unique and there should be no issues involving the untameable growth of gradients.

This paper is organised as follows. Starting from the $(4+1)$-dimensional Chern-Simons action for a free Abelian vector potential (connection) $A=A_\mu \mathrm{d}x^\mu$ (with $\mu=0,1,2,3,4$) in Section \ref{sec:free}, it is shown that the field strength (curvature two-form) $\mathrm{F}=\mathrm{d}A$ necessarily possesses at least three independent null eigenvectors which can be used to define an advecting velocity field. The resulting Euler-Lagrange equations become advection equations, where one can identify prototypical forms of helicity and entropy conservation. The discussion is then extended to the interacting case in which the connection is minimally coupled to an external current in Section \ref{sec:J^mu}. In this new setting that the external current is seen as a null eigenvector of $\mathrm{F}$, that is necessarily conserved, which guarantees gauge invariance of the action and switches on propagating degrees of freedom excited by a charge density that has the structure of the Rossby-Ertel's potential vorticity. The main results are presented in Section \ref{sec:flu_dyn}, where the external current is identified with the Chern-Simons field, in a non-relativistic setting, introducing a ``dressing'' gauge field. The resulting Euler-Lagrange equations of this self-interacting theory are exactly the dynamical equations for a dissipationless compressible fluid in $(3+1)$ dimensions, but with a constraint on the initial profile of the Rossby-Ertel's potential vorticity, whose implications are explored. In this Section, the tools provided by the Chern-Simons formulation are employed to extend Moffatt's helicity invariant for baroclinic fluids to the case of arbitrary boundary conditions, clarifying the role of baroclinity in helicity production. Examples based on an ideal fluid in rotating scenarios are given, including exact solutions and show how the present formulation relates to Crocco's equation and Kuznetsov's Lorentz-force hydrodynamics. Finally, Section \ref{sec:conclusions} contains concluding remarks.

In this paper we use the following convention regarding tensor indices and tensor dimensions: 

\begin{itemize}
\item Greek indices ($\alpha, \beta, \mu, \nu, \ldots$) range from $0$ to $4$. The corresponding tensors are understood as fully $(4+1)$-dimensional.
\item Latin indices ($i, j, k, l, \ldots$) range from $1$ to $4$. The corresponding tensors are understood as $4$-dimensional.
\item Restricted Latin indices ($a, b, c, d, \ldots$) range from $1$ to $3$. The corresponding tensors are understood as $3$-dimensional, and we will usually denote them using \textbf{boldface} characters.
{\item Hatted Greek indices ($\hat{\alpha}, \hat{\beta}, \hat{\mu}, \hat{\nu}, \ldots$) range from $0$ to $3$. The corresponding tensors are fully $(3+1)$-dimensional.}
\item Only when necessary, a tensor's dimension will be explicitly shown to avoid confusion. In such a case, if a tensor $V$ is $D$-dimensional, it will be denoted $V_{(D)}$ or $V^{(D)}$.
\end{itemize}

%%%%%%%%%%%%%%%%%%%%%%%%%%%%%%%%%%%%%%%%%%%%%%%
\section{Free $(4+1)${-dimensional} Chern-Simons theory}
%%%%%%%%%%%%%%%%%%%%%%%%%%%%%%%%%%%%%%%%%%%%%%%
\label{sec:free}

Consider the $5$-dimensional Chern-Simons action for a non-interacting Abelian connection $A$,
\begin{equation}
I_0[A] = \int_{{\mathcal{M}}_5} \frac{1}{3!} A\wedge \mathrm{F}\wedge \mathrm{F}\, .
\end{equation}
The field strength (curvature) is $\mathrm{F}=\mathrm{d}A$, and the {Euler-Lagrange} equations, 
\begin{equation}
    \mathrm{F} \wedge \mathrm{F}=0\,,\label{F^2}
\end{equation}
obtained by varying the above action with respect to $A$, can also be explicitly written as (sum over repeated indices is understood)
\begin{equation} \label{FF=0}
\epsilon^{\alpha \beta \mu \nu \lambda}\, \mathrm{F}_{\alpha \beta}\,\mathrm{F}_{\mu \nu}=0 \,, \qquad \lambda = 0, \ldots, 4\,.
\end{equation}
%%%%%%%%%%%%%%%%%%%%%%%%%%%%%%%%%%%%%%%%%%%%%%%
\subsection{Five-dimensional analysis}
%%%%%%%%%%%%%%%%%%%%%%%%%%%%%%%%%%%%%%%%%%%%%%%
{For $\lambda=i, 0$ equations \eqref{FF=0} take the form 
\begin{align} \label{Fk0}
\epsilon^{ijkl}\mathrm{F}_{jk}\,\mathrm{F}_{l0} &=0\,,\\ \label{Fij}
\epsilon^{ijkl}\,\mathrm{F}_{ij}\,\mathrm{F}_{kl} &=0 \,,
\end{align}
respectively, where $\epsilon^{j k l m}$ is the totally antisymmetric tensor in $4$-dimensional {Euclidean} space.
In an explicit matrix notation, equation \eqref{Fk0} reads:
\begin{equation}
\label{eq:CS_matrix}
\left(\begin{array}{c}
0\\
\mathrm{F}_{34}\\
-\mathrm{F}_{24}\\
\mathrm{F}_{23}
\end{array}  \begin{array}{c}
-\mathrm{F}_{34}\\
0\\
\mathrm{F}_{14}\\
-\mathrm{F}_{13}
\end{array} \begin{array}{c}
\mathrm{F}_{24}\\
-\mathrm{F}_{14}\\
0\\
\mathrm{F}_{12}
\end{array}  \begin{array}{c}
-\mathrm{F}_{23}\\
\mathrm{F}_{13}\\
-\mathrm{F}_{12}\\
0
\end{array}
\right) \left(\begin{array}{c}
\mathrm{F}_{01}\\
\mathrm{F}_{02}\\
\mathrm{F}_{03}\\
\mathrm{F}_{04}
\end{array}
\right) = \left(\begin{array}{c}
0\\
0\\
0\\
0
\end{array}
\right)\,,
\end{equation}
where the column vector $\mathrm{F}_{0i}$ is a null vector of the $4\times 4$ matrix $\epsilon^{ijkl}\mathrm{F}_{kl}$.
Equations \eqref{eq:CS_matrix} involve time derivatives of $A_k$ through $\mathrm{F}_{0k}$. These equations are not linearly independent because the $4 \times 4$ tensor $\mathrm{F}_{lm}$ has vanishing determinant, with a double zero: $\det[\mathrm{F}]= (\epsilon^{ijkl}\mathrm{F}_{ij}\,\mathrm{F}_{kl})^2$, by virtue of (\ref{Fij}). Hence, the rank of of $\mathrm{F}$ is 2 or 0 and equation \eqref{Fij} is a constraint --consistent with the other equations-- which implies that this system can have at most two nontrivial propagating solutions in the space of fields. This means that, generically, the column vector $\mathrm{F}_{0i}$ that solves (\ref{eq:CS_matrix}) belongs to a $2$-dimensional subspace and (\ref{eq:CS_matrix}) represents $2$ independent equations for $2$ unknowns at each spacetime point. It turns out that these two non vanishing degrees of freedom can be eliminated by gauge fixing, and detailed Hamiltonian analysis shows that this system has no propagating local degrees of freedom \citep{banados1996existence}. A further degeneracy occurs   
at spacetime points where $\mathrm{F}_{jk} = 0$ for all $j, k = 1, \ldots, 4.$ There, the double zero of the determinant becomes a quadruple zero. In section \ref{sec:F4} we will see that the spacetime points where this happens are conserved if one assumes differentiability of the relevant fields.

%%%%%%%%%%%%%%%%%%%%%%%%%%%%%%%%%%%%%%%%%%%%%%%%%
\subsection{Explicit solution for $\mathrm{F}_{0j}$} % 2.2%
%%%%%%%%%%%%%%%%%%%%%%%%%%%%%%%%%%%%%%%%%%%%%%%%%%

Equation (\ref{eq:CS_matrix}) can be solved for the vector $(\mathrm{F}_{01},\mathrm{F}_{02},\mathrm{F}_{03},\mathrm{F}_{04})^T$ as a linear combination of the null-space eigenvectors of the matrix in that equation, as follows:
\begin{equation}
\label{eq:CS_sol}
\left(\begin{array}{c}
\mathrm{F}_{01}\\
\mathrm{F}_{02}\\
\mathrm{F}_{03}\\
\mathrm{F}_{04}
\end{array}
\right) = V^1 \left(\begin{array}{c}
0\\
-\mathrm{F}_{12}\\
-\mathrm{F}_{13}\\
-\mathrm{F}_{14}
\end{array}
\right)
+
V^2 \left(\begin{array}{c}
\mathrm{F}_{12}\\
0\\
-\mathrm{F}_{23}\\
-\mathrm{F}_{24}
\end{array}
\right)
+
V^3 \left(\begin{array}{c}
\mathrm{F}_{13}\\
\mathrm{F}_{23}\\
0\\
-\mathrm{F}_{34}
\end{array}
\right)
+
V^4 \left(\begin{array}{c}
\mathrm{F}_{14}\\
\mathrm{F}_{24}\\
\mathrm{F}_{34}\\
0
\end{array}
\right)\,,
\end{equation}
where $(V^1, V^2, V^3, V^4)$ are four arbitrary functions {in the five-dimensional} spacetime. These four functions can be interpreted as the four {spatial} components of an arbitrary five-vector field of components 
\begin{equation}
\label{eq:def_V}
V^\mu = (1,V^1, V^2, V^3, V^4)\,.
\end{equation}
Expression (\ref{eq:CS_sol}) {can also be written as 
\begin{equation}
\label{eq:null_5d}
V^\mu \mathrm{F}_{\mu \nu} = 0\,, \quad \nu = 0, \ldots , 4\,,
\end{equation}
which states that $V^{\mu}$ is a null vector of $\mathrm{F}_{\mu \nu}$.} 
Notice that the $\nu=0$ component of this equation is a consequence of the other 4 components, and the latter are a restatement of equation (\ref{eq:CS_sol}). 

The vector $V^\mu$ is not unique since it can be multiplied by  an arbitrary function $\rho$ and therefore, for a given $\mathrm{F}_{\mu \nu}$ one can also write $\tilde{V}^\mu \mathrm{F}_{\mu \nu} = 0$ with $\tilde{V} = \rho(1,V^1, V^2, V^3, V^4)$. In summary, because of these invariances the null eigenvector $\tilde{V}$ has five  components, of which only 2 are independent plus an overall rescaling (see Appendix \ref{eigen} where the general null eigenvector solution to eq. \eqref{eq:null_5d} is discussed).  

Although the column vectors in (\ref{eq:CS_sol}) span a $2$-dimensional subspace at each spacetime point, it is useful to write the solution in the equivalent redundant forms (\ref{eq:CS_sol}) or (\ref{eq:null_5d}) for two reasons: (i) It allows for a global approach in terms of the choice of basis vectors of the corresponding $2$-dimensional solution subspace: generically, at each point of the spacetime, one can always find two basis vectors amongst the four vectors in the expansion; (ii) It gives rise to a very simple interpretation of the equations in terms of advection equations along the flow vector field $V$, which we discuss now and which will carry on to theories that include interaction with an external source and self-interaction.

%%%%%%%%%%%%%%%%%%%%%%%%%%%%%%%%%%%%%%%%%%%%%%%%%%%%%%%%%%%%%%%%%%
\subsection{Evolution of $A_\mu$ as an advection equation} % 2.3 %
%%%%%%%%%%%%%%%%%%%%%%%%%%%%%%%%%%%%%%%%%%%%%%%%%%%%%%%%%%%%%%%%%%

Equation (\ref{eq:null_5d}) can be written in the coordinate-free form
$i_V \mathrm{d}A = 0$, {where $V$ is given in \eqref{eq:def_V}.} Assuming $V$ to be differentiable, Cartan's formula for the Lie derivative along a vector field $V$, $\mathcal{L}_V = i_V \mathrm{d} + \mathrm{d} i_V$, allows to rewrite (\ref{eq:null_5d}) as advection equation:
\begin{equation}
\label{eq:Lie_A}
\mathcal{L}_V A = \mathrm{d} (i_V A)\,.
\end{equation}

As explained after equation (\ref{eq:null_5d}), the $0$-component of this equation is not independent, so we will consider its spatial components only. Noting $(\mathcal{L}_{V}A)_\mu = V^\nu \partial_\nu A_\mu+A_\nu \partial_\mu V^\nu$ and using the special form of $V$ given in equation (\ref{eq:def_V}), the `spatial' components involve the dynamic equations: 
\begin{equation}
\label{eq:CS_A_mu}
\left(\partial_0 + V^j \partial_j\right) A_k = \partial_k A_0 + V^j \partial_k A_j\,, \qquad k =1, \dots, 4\,.
\end{equation}
It is clear that the $4$ components $A_k$ are dynamical whereas $A_0$ is not, as in electromagnetism. A coordinate-free form of this equation is obtained in terms of the Lie derivative along the vector field $V_{(4)} = V^j \partial_j$ of the $4$-dimensional $1$-form ${\mathbb{A}}^{(4)} := A_j \mathrm{d}x^j$, giving
\begin{equation}
\label{eq:Lie}
\left(\partial_0 + \mathcal{L}_{V_{(4)}}\right) \mathbb{A}^{(4)} = \mathrm{d}_{(4)} \left(A_0 + i_{V_{(4)}}{\mathbb{A}^{(4)}}\right)\,,
\end{equation}
where the label ``$(4)$'' refers to the `spatial' $4$-dimensional restrictions. 

%%%%%%%%%%%%%%%%%%%%%%%%%%%%%%%%%%%%%%%%%%%%%%%%%%%%%%%%%%%%%%%%%%%%%%
\subsection{Evolution of $\,\mathbb{F}^{(4)}\,$ and conservation of its zeroes} % 2.4%
%%%%%%%%%%%%%%%%%%%%%%%%%%%%%%%%%%%%%%%%%%%%%%%%%%%%%%%%%%%%%%%%%%%%%%
\label{sec:F4}

Taking the $4$-dimensional exterior derivative of equation (\ref{eq:Lie}) we obtain
\begin{equation}
\label{eq:Lie_F}
\left(\partial_0 + {\mathcal{L}}_{V_{(4)}}\right) {\mathbb{F}^{(4)}} = 0\,,
\end{equation}
where $\mathbb{F}^{(4)}=\mathrm{d}_{(4)}{\mathbb{A}^{(4)}}=\partial_i A_j dx^i \wedge dx^j$ and $\mathcal{L}_{V}\mathrm{F}_{kl} := V^j \partial_j \mathrm{F}_{kl}+ \partial_k V^j \mathrm{F}_{jl}-\partial_l V^j \mathrm{F}_{jk}$. Equation (\ref{eq:Lie_F}) means that the tensor $\mathbb{F}^{(4)}$ is `conserved' along the flow defined by the vector field $V$. In particular, the regions of spacetime where $\mathbb{F}^{(4)}=0$ move with the flow along the vector field $V$. This follows from the fact that
$$\frac{1}{4}\left(\partial_0 + V^l \partial_l\right) \sum_{j,k=1}^4\left(\mathrm{F}_{jk}\mathrm{F}_{jk}\right) = \sum_{j,l,k=1}^4 \mathrm{F}_{jl}V^l_{,k}\mathrm{F}_{kj}\,,$$
and the RHS vanishes at the zeroes of $\mathrm{F}$, so the zeroes are ``conserved'' by the flow: they cannot be created or destroyed and and move with the flow. Therefore, we conclude that in a differentiable flow the degenerate condition that the matrix in (\ref{eq:CS_matrix}) has rank zero, corresponds to these conserved zeroes.

%%%%%%%%%%%%%%%%%%%%%%%%%%%%%%%%%%%%%%%%%%%%%%%%%%%%%
\subsection{Reduction to $3 + 1$ dimensions} 
\label{sec:3+1} % 2.5 %
%%%%%%%%%%%%%%%%%%%%%%%%%%%%%%%%%%%%%%%%%%%%%%%%%%%%%
Let us assume the five-dimensional spacetime $\mathcal{M}_5$ to be the product of a four-dimensional Lorentzian spacetime $\mathbb{R}^{1,3}$ and a compact factor $S$. Using coordinates $x^{\hat{\mu}}=(x^0, x^a) \in \mathbb{R}^{1,3}$ and $x^4  \in S$, the Abelian connection reads  
\begin{align} \nonumber
A=&A_{\hat{\mu}}\mathrm{d}x^{\hat{\mu}} + A_4 \mathrm{d}x^4 \\
\label{A-reduced}
=& A_0 \mathrm{d}x^0 + A_a \mathrm{d}x^a + A_4 \mathrm{d}x^4\,.
\end{align}
\
We further assume $S$ to be a compact isometry orbit in $\mathcal{M}_5$ (e.g., $S^1$). Since the coordinate $x^4$ along $S$ parameterises the orbit of a Killing vector field (a global isometry), the simplest dynamical excitations are independent of $x^4$, and {in this case} equations \eqref{F^2} become
\begin{subequations}
  \label{phidot-B}
    \begin{empheq}[left={{\epsilon^{\hat \mu\hat\nu\hat\rho\hat\sigma}\partial_{\hat\nu}A_4 \mathrm{F}_{\hat\rho\hat\sigma}=0} \quad \Rightarrow \quad \empheqlbrace\,}]{align}
      \mathbf{B} \cdot \nabla A_4 &= 0\,,
        \label{B.Del phi} \\
      \mathbf{E}\times \nabla A_4 - \mathbf{B} \partial_0{A_4} &= \mathbf{0},
        \label{ExDel phi}
    \end{empheq}
\end{subequations}

\begin{eqnarray}
\label{EB}
\quad{\epsilon^{\hat \mu\hat\nu\hat\rho\hat\sigma}\mathrm{F}_{\hat\mu\hat\nu}\mathrm{F}_{\hat\rho\hat\sigma}=0} \quad \Rightarrow &\qquad &\qquad\qquad -\mathbf{E}\cdot\mathbf{B}=0 \, , 
\end{eqnarray}
where we have used the standard ``electric" and ``magnetic" fields, $\mathbf{E}=-\mathrm{F}_{0a}\mathbf{\hat{x}}^a$ and $\mathbf{B}=\frac{1}{2}\epsilon_{abc}\mathrm{F}_{ab}\mathbf{\hat{x}}^c$, which satisfy the Bianchi identities $\mathrm{d F}=0$, namely $\partial_0 \bt{B} + \nabla \times \bt{E} = 0$ and $\nabla \cdot \bt{B}$. 
Note that in spacetime regions where $\partial_0{ A_4} \neq 0$, Eqs.~(\ref{EB}) and \eqref{B.Del phi} follow from (\ref{ExDel phi}).

The null-vector equations (\ref{eq:null_5d}) become, after writing $V^\mu = (1,\bt{V},V^4)$ where $\bt{V}$ is a $3$-dimensional vector field,
\begin{eqnarray}
\label{eq:null_split_0}
\bt{V} \cdot \bt{E} - V^4 \partial_0{ A_4} &=& 0\,,\\
\label{eq:null_split_4}
\partial_0{ A_4} + \bt{V} \cdot\nabla A_4 &=& 0\,,\\
\label{eq:null_split_a}
{V^4\nabla  A_4}+ (\bt{E} + \bt{V} \times \bt{B}) &=& \bt{0}\,.
\end{eqnarray}
In summary, the Euler-Lagrange equations $\mathrm{F}\wedge \mathrm{F}=0$, equivalently written as the five equations \eqref{phidot-B}--\eqref{EB}, imply the five null-vector equations \eqref{eq:null_split_0}--\eqref{eq:null_split_a}. Alternatively, the set of five equations given by (\ref{eq:null_split_4}), (\ref{eq:null_split_a}) and \eqref{B.Del phi}, form a fundamental set, equivalent to \eqref{phidot-B}--\eqref{EB}, as they are independent and imply the remaining equations: (\ref{EB}), (\ref{eq:null_split_0}) and (\ref{ExDel phi}).

%%%%%%%%%%%%%%%%%%%%%%%%%%%%%%%%%%%%%%%%%%%%%%%%%%%%%%%%%%%%
\subsection{Advection formulation in the $(3+1)$ reduction}
\label{sec:adv}

%%%%%%%%%%%%%%%%%%%%%%%%%%%%%%%%%%%%%%%%%%%%%%%%%%%%%%%%%%%%
It is useful to write the fundamental set (\ref{eq:null_split_4}), (\ref{eq:null_split_a}), \eqref{B.Del phi} in terms of Lie derivatives and the advection equation (\ref{eq:Lie}) as
\begin{eqnarray}
\label{eq:3d_red_constraint}
{\mathbf{B}}\cdot {\nabla}A_4 &=& 0\,,\\ 
\label{eq:3d_red_A4}
(\partial_0 + \mathcal{L}_{\bt{V}})A_4 &=& 0\,,\\
\label{eq:3d_red_advection}
(\partial_0 + \mathcal{L}_{\bt{V}})\bt{A} &=& \nabla (A_0 + V^b A_b) + V^4 \nabla A_4\,,
\end{eqnarray}
where $\bt{A} = A_a \mathrm{d}x^a$ is the $3$-dimensional vector potential and the Lie derivatives $\mathcal{L}_{\bt{V}}A_4$ and $\mathcal{L}_{\bt{V}}\bt{A}$ are given in Appendix \ref{sec:Lie_der}. The advection equations \eqref{eq:3d_red_A4} and \eqref{eq:3d_red_advection} determine the evolution of the scalar $A_4$ and the vector potential $\bt{A}$, as they are advected by $\bt{V}$. In the following, we shall refer to $\bt{V}$ defined in \eqref{eq:def_V} as the flow velocity. In contrast, \eqref{eq:3d_red_constraint} is a \emph{constraint}, which unsurprisingly is preserved by this evolution.

 Recalling that $\bt{A}$ and $\bt{B}$ are related by ${(}\bt{d} \bt{A}{)}_{a b} = \epsilon_{abc} B^c$,
we take the $3$-dimensional exterior derivative $\bt{d}$ of equation (\ref{eq:3d_red_advection}) to get
$(\partial_0 + \mathcal{L}_{\bt{V}})\bt{d} \bt{A} = \bt{d} V^4 \wedge \bt{d}A_4$,
or in terms of the vector field $\bt{B}$, 
\begin{equation}
\label{eq:3d_red_B}
    \partial_0 \bt{B} - \nabla \times (\bt{V}\times \bt{B})= \nabla V^4 \times \nabla A_4\,.
\end{equation}

{The main advantage of the advection formulation is that one can solve the Euler-Lagrange equations by simply integrating forward in time the Cauchy problem based on \eqref{eq:3d_red_A4}--\eqref{eq:3d_red_advection}, with the only proviso that the constraint \eqref{eq:3d_red_constraint} must be satisfied initially. Strategies for finding initial conditions that are compatible with the constraint \eqref{eq:3d_red_constraint} can vary. For example, a challenging strategy is to assume a starting field $\bt{A}$ which then gives an initial magnetic field $\bt{B} = \nabla \times \bt{A}$, so the constraint \eqref{eq:3d_red_constraint} implies that the starting scalar field $A_4$ is constant along the magnetic field lines. This strategy is challenging because it is not uncommon to find that magnetic field lines are chaotic (see for example \cite{hosoda2009ubiquity}), which would not allow for a continuous field $A_4$ to exist. A safer strategy is to start with a given non-uniform smooth scalar field $A_4$ and $\bt{A} = 0$, which gives $\bt{B}=0$ initially, satisfying the constraint. According to the evolution \eqref{eq:3d_red_B}, the magnetic field will quickly pick up non-zero values at subsequent times, while the constraint \eqref{eq:3d_red_constraint} continues to hold. Incidentally, this last example illustrates the ``conservation of zeroes'' result of section \ref{sec:F4}: initially, both $\bt{B}$ and $\nabla A_4$ are zero at the extreme points of $A_4$ (that is, the spatial components $\mathrm{F}_{ij}$ vanish at these points), and therefore these simultaneous zeros of $\bt{B}$ and $\nabla A_4$ will be advected by the flow velocity $\bt{V}$ at later times, despite the fact that $\bt{B}$ can develop non-zero values everywhere else.}

%%%%%%%%%%%%%%%%%%%%%%%%%%%%%%%%%%%%%%%%%%%%%%
\subsection{The proto-helicity: a conserved current of the advection formulation}
\label{sec:pHelicityCurrent}
%%%%%%%%%%%%%%%%%%%%%%%%%%%%%%%%%%%%%%%%%%%%%%
Based on our previous discussion, we now construct some conserved quantities characteristic of the flow.
In particular, we look for conserved quantities that depend on $\bt{A}$, $A_4$ and their derivatives. To begin with, 
let us derive from the field equations \eqref{phidot-B}--\eqref{EB} an important conserved current, which we will call the \emph{proto-helicity current density}, {as it will be clarified at the end of this {preamble}.}
Still within our reduction to $3 + 1$ dimensions, namely assuming that no field depends on $x^4$, we define
\begin{equation}
\label{eq:j_heli_def}
{{\heli}^{\hat\mu} := \frac{1}{2}\,\epsilon^{\hat\mu\hat\nu\hat\rho\hat\sigma} A_{\hat\nu}\mathrm{F}_{\hat\rho\hat\sigma}\,, \qquad \hat\mu = 0, \ldots, 3\,, \qquad {\heli}^{4} := 0\,.}
\end{equation}
{Explicitly we have, in terms of the fields $A_\mu, \bt{B}$ and $\bt{E}$,
\begin{equation}
    \label{eq:j_heli_def_2}
{\heli}^{\mu} = (\heli^0, \bt{\heli},0), \quad \text{with} \quad  \heli^0 = A_a B^a,  \quad \bt{\heli} = - A_0\,{\bf B} - {\bf A}\times {\bf E}\,.
\end{equation}}
The divergence of this current reads:
\begin{equation}
\label{eq:j_heli_div}
    {\partial_{\mu}{\heli}^{\mu} =  }  \partial_{\hat\mu}{\heli}^{\hat\mu}=\frac{1}{2}\,\epsilon^{\hat\mu\hat\nu\hat\rho\hat\sigma} \partial_{\hat\mu}A_{\hat\nu}\mathrm{F}_{\hat\rho\hat\sigma}\,.
\end{equation}
The right-hand side vanishes {by equations \eqref{phidot-B}--\eqref{EB}} and we find the following on-shell conservation law:
\begin{equation}
\label{conservation_law_PE}
{ \partial_{\mu}{\heli}^{\mu}=0\, \qquad \text{or} \qquad \partial_0 (A_a B^a) + \nabla \cdot \left(-A_0\,{\bf B}-{\bf A}\times {\bf E}\right) = 0\,.}
\end{equation}

Now, suppose ${\mathcal{M}}_3$ is a spatial slice at time $t$ of the four-dimensional Lorentzian spacetime, and consider a subset $\Omega(t) \subset {\mathcal{M}}_3$, which is advected by the flow velocity $\bt{V}$. Define the conserved quantity corresponding to equation \eqref{conservation_law_PE} as the \emph{proto-helicity charge}:}
\begin{equation}
\label{eq:proto-helicity}
    \mathcal{H}(t):=\int_{\Omega(t)}\,{\heli}^0\,\mathrm{d}^3x =  {\int_{\Omega(t)}\,A_a B^a\,\mathrm{d}^3x =   \int_{\Omega(t)} \bt{A} \wedge \bt{d} \bt{A}}\,.
\end{equation}
{This is the Chern-Simons $3$-form, also known as the Hopf invariant. Historically, helicity has played an important role in understanding topological aspects of flow configurations, particularly regarding the topology of vortex lines in hydrodynamics and the topology of magnetic field lines in magnetohydrodynamics, and their applications to dynamo excitations, turbulent cascades, and coherent structures \citep{doi:10.1073/pnas.1400277111,moffatt1992helicity,ricca2012introduction}.  More generally, {{expression \eqref{eq:proto-helicity} represents a coupling between $\bt{A}$ and a two-dimensional membrane \citep{mivskovic2009couplings}, which occurs naturally at the boundary of topological insulators, where the two-dimensional surface of the material traces a three-dimensional surface evolving in time \citep{huerta2012optical}.}}  Notice that, at this stage, our proto-helicity does not depend on the velocity field; rather, it depends on the advected field only, hence the prefix ``proto''. In subsequent sections we will relate dynamically the velocity field to the advected field, and a full picture of a fluid's helicity will emerge.}

 {\subsubsection{Advection and conservation of the proto-helicity}}

Let us note the following  property of  {${\heli}^{\mu}$}:
\begin{equation}
    {\heli}^{\hat\mu}\,\mathrm{F}_{\hat{\mu}\nu}=0\,, \qquad \nu = 0, \ldots, 4\,,\label{JF}
\end{equation}
which follows from \eqref{ExDel phi} and \eqref{EB}. {This means that $h^\mu$ is a null eigenvector of $\mathrm{F}_{\mu\nu}$, in the free Chern-Simons theory.}
Here we emphasize the difference between ${\heli}^{\hat\mu}$ and the  {flow} velocity 4-vector $V^{\hat{\mu}}$, which satisfies, instead,  equation  {\eqref{eq:null_5d}, that is}:
\begin{equation}V^{\hat\mu}\,\mathrm{F}_{\hat{\mu}\nu}+V^4\,\mathrm{F}_{4\nu}=0\,, \qquad \nu = 0, \ldots, 4\,.\label{VF}
\end{equation}
Since $V^4 \neq 0$ in general, we conclude that the  {\emph{apparent velocity}} ${\bf V}'=(V^{\prime a})$ at which the proto-helicity  {charge} $\mathcal{H}$ is transported, defined as
\begin{equation}
V^{\prime a} := \frac{{\heli}^a}{{\heli}^0}=-\frac{1}{A_bB^b}\,\left(A_0\,{\bf B}+{\bf A}\times {\bf E}\right)\,,
\end{equation}
may not coincide with the  {flow} velocity ${\bf V}=(V^a)$. {The latter, we recall, is not uniquely defined in the free $5$-dimensional Chern-Simons theory. } Interestingly though, we can compare these two velocities: from equation \eqref{eq:null_split_a} we deduce ${\bf A}\times {\bf E} = - V^4 {\bf A}\times {\nabla  A_4} - {\bf A}\times (\bt{V} \times \bt{B})$, so 
\begin{equation}
    \label{eq:velo_diff}
\heli^0 ({\bf V}' - {\bf V}) = - \left(A_0\,{\bf B}+{\bf A}\times {\bf E} + A_bB^b  {\bf V} \right) = -  \left((A_0+A_b V^b)\,{\bf B} - V^4 {\bf A}\times {\nabla  A_4} \right)\,.
\end{equation}
Due to this relation, we can re-write the conservation law \eqref{conservation_law_PE} in the advection form
\begin{equation}
    \label{eq:conservation_law_PE_2}
    \partial_0 \heli^0 + \nabla \cdot (\heli^0 \bt{V}) = \nabla \cdot \left((A_0+A_b V^b)\,{\bf B} - V^4 {\bf A}\times {\nabla  A_4} \right)\,.
\end{equation}
Now, from here we can revert to an integral formulation. The time derivative of the proto-helicity charge defined in equation \eqref{eq:proto-helicity} is, using the Leibniz-Reynolds transport theorem,
$$\frac{\mathrm{d}}{\mathrm{d} t} \left(\int_{\Omega(t)}\,{\heli}^0\,\mathrm{d}^3x\right) = \int_{\Omega(t)}\,\left(\partial_0 \heli^0 + \nabla \cdot (\heli^0 \bt{V})\right)\,\mathrm{d}^3x\,,$$
and using equation \eqref{eq:conservation_law_PE_2} and  the divergence theorem, this gives a ``flux'' boundary integral:
\begin{equation}
    \label{eq:heli_flux}
\frac{\mathrm{d}}{\mathrm{d} t} \left(\int_{\Omega(t)}\,{\heli}^0\,\mathrm{d}^3x\right) = \oint_{\partial\Omega(t)}\,  \,\left((A_0+A_b V^b)\,{\bf B} - V^4 {\bf A}\times {\nabla  A_4} \right)\cdot\bt{n}\,\mathrm{d}^2x\,,
\end{equation}
where $\bt{n}$ is the outward-directed normal to $\partial\Omega(t)$ and $\mathrm{d}^2x$ the surface element on it. In order for this flux term to be equal to zero, one has two strategies: (i) To set ${\bf V} = {\bf V}'$ (equivalently, $A_0+A_b V^b=V^4=0$) everywhere (too restrictive). (ii) To define $\Omega(t)$ such that $\partial \Omega(t)$ is an isosurface of the scalar $A_4$, which is consistent with the conservation of $A_4$, equation \eqref{eq:3d_red_A4}. Then, $\bt{n} \propto \nabla A_4$, hence, using the constraint \eqref{eq:3d_red_constraint}, it follows that the right hand side of   equation \eqref{eq:heli_flux} is equal to zero at all times, and conservation of proto-helicity charge is proved.\\
We provide, in Appendix \ref{app:heli_covariant}, an alternative derivation (elegant and covariant) of these results.

%%%%%%%%%%%%%%%%%%%%%%%%%%%%%%%%%%%%%%%%%%%%%%%%%%%%%%%%%%
\section{Interaction with an external current density $J^\mu$}
\label{sec:J^mu}
%%%%%%%%%%%%%%%%%%%%%%%%%%%%%%%%%%%%%%%%%%%%%%%%%%%%%%%%%%
The previous discussion applies to an isolated dynamical system. We now introduce the interaction of the connection $A$ with the environment characterised by an external source $J$. Thus, consider the functional
\begin{equation}\label{int-action}
I_1[A;J] = \int_{{\mathcal{M}}_5} \left[ {\frac{1}{3!}} A\wedge \mathrm{F}\wedge \mathrm{F} -  A_\mu {J}^\mu \mathrm{d}^5x\right]= \int_{{\mathcal{M}}_5} \left[ {\frac{1}{3!}} A\wedge \mathrm{F}\wedge \mathrm{F} + A\wedge {}^*J\right]\,.
\end{equation}
The Euler-Lagrange equations obtained after varying with respect to $A$ read
\begin{eqnarray}
\label{eq:i1}
 {\frac{1}{2}} \mathrm{F}\wedge \mathrm{F} &=& {-} {}^*{J}\,,
\end{eqnarray}
where ${}^*$ denotes the Hodge dual operator in five dimensions{, whose precise definition is given in Appendix \ref{notations metric}}. From these equations we can draw two results. First, {taking the exterior derivative of Eq.~(\ref{eq:i1}),
and} applying the {Bianchi identity $\mathrm{d}\mathrm{F} \equiv 0$,} %
%\mathrm{d} (F\wedge F) = 0
%to
leads to {$\mathrm{d}{}^*J=0$, that is}
\begin{equation}
\label{eq:J_5d_cons}
\partial_\mu J^\mu = 0\,.
\end{equation}
So the external current $J$ is necessarily conserved, as in electrodynamics. Second, using the identity
$ \epsilon^{\mu \nu \lambda \rho \sigma}\mathrm{F}_{\nu \lambda}\mathrm{F}_{\rho \sigma}\mathrm{F}_{\mu \alpha}\equiv 0$
in Eq.~(\ref{eq:i1}), we obtain that $J$ must be a null eigenvector of $\mathrm{F}$: %[we recover \eqref{eq:null_5d}] 
\begin{equation}
\label{eq:J_null_5d}
J^\mu \mathrm{F}_{\mu \nu} = 0\,, \quad \nu=0, \ldots, 4\,.
\end{equation}
{In other words, while in the free case the dynamical equations implied the existence of at least one null eigenvector $V^\mu$, in the interacting case the external source turns out to be such a null vector. Although both $V^\mu$ and $J^\mu$ are null vectors of $\mathrm{F}_{\mu \nu}$, there is an important difference between them, related to the fact that $J^\mu$ is required by consistency to be a conserved current, which means that one cannot arbitrarily set its zeroth component equal to one, as it was the case with $V^\mu$. {{Since equation (\ref{eq:J_null_5d}) is formally the same as \eqref{eq:null_5d},} we re-use the notation of the free case and decompose the current in terms of a spatial  {charge} density $\rho := J^0$ and a vector field $V$ which can be interpreted as a velocity field:
\begin{equation}
\label{eq:decomp_J^mu}
    J^\mu = \rho V^\mu, \qquad V^\mu = (1, V^1, V^2, V^3, V^4) =: (1, \bt{V}, V^4) \,.
\end{equation}}
}

{Equation \eqref{eq:J_null_5d} has an interesting physical interpretation as well. It describes a force-free plasma, that is, a situation in which there is no exchange of energy and momentum between the electromagnetic field and the particles that make up the source. This feature, when restricted to the (3+1)-dimensional setting, is a relevant ingredient in the description of the dynamics of plasmas around compact astrophysical objects such as pulsars and black holes \citep{Brennan:2013kea}. {Also, as we will show in section \ref{sec:Crocco}, as a fluid equation this is equivalent to Crocco's equation \citep[Ch.3]{currie2002fundamental} and is related to Kuznetsov's Lorentz-force law hydrodynamics \citep{kuznetsov2008mixed}.}}

As in the free case, the equations of motion have an advection formulation.  {A system of transport equations is obtained, this time in terms of a \emph{particular} {(uniquely defined modulo a  rescaling)} null eigenvector.} The novel feature of the interacting case is that, having a \emph{given} current $J$ in the RHS of equation \eqref{eq:i1}, breaks the degeneracy of the tensor $\mathrm{F}$ that was found in the free case. {In particular, for $J^0=0$ Eq.\eqref{eq:i1} implies that the field $A_\mu$ has exactly zero locally propagating degrees of freedom \citep{banados1996existence}. This means that the case $J^0 =0$ is qualitatively different from the situation in which $J^0$ is very small but nonzero.} {The sole presence of a non vanishing $J^0$ switches on propagating degrees of freedom of the field $A_\mu$. This contradicts the perturbative expectation in which the dynamics of the interacting case is continuously connected to that of the free system. Here the free case is not necessarily a good approximation for the slightly interacting case, no matter how small $J^0$ would be.}

%%%%%%%%%%%%%%%%%%%%%%%%%%%%%%%%%%%%%%%%%%%%%%%%%%%%%%%%%%%%
\subsection{Reduction to $3+1$ dimensions with $J = \rho(1, \bt{V}, V^4)$}
%%%%%%%%%%%%%%%%%%%%%%%%%%%%%%%%%%%%%%%%%%%%%%%%%%%%%%%%%%%%
{Assuming the form \eqref{A-reduced} for the reduction} to $3+1$ dimensions, the equations of motion \eqref{eq:i1} read

\begin{subequations}
  \label{phidot-B_J}
    \begin{empheq}[left={{ {\frac{1}{2}}\, \epsilon^{\hat \mu\hat\nu\hat\rho\hat\sigma}\partial_{\hat\nu}A_4 \mathrm{F}_{\hat\rho\hat\sigma}=J^{\hat\mu}} \quad \Rightarrow \quad \empheqlbrace\,}]{align}
      \mathbf{B} \cdot \nabla A_4 &= {J^0}\,,
        \label{B.Del phi_J} \\
      \mathbf{E}\times \nabla A_4 - \mathbf{B} \partial_0{A_4} &= {\mbf{J}},
        \label{ExDel phi_J}
    \end{empheq}
\end{subequations}
\begin{eqnarray}
\label{EB_J}
\quad{{\textstyle  {\frac{1}{8}}}\,\epsilon^{\hat \mu\hat\nu\hat\rho\hat\sigma}\mathrm{F}_{\hat\mu\hat\nu}\mathrm{F}_{\hat\rho\hat\sigma}={J^4}} \quad \Rightarrow &\qquad &\qquad\qquad -\mathbf{E}\cdot\mathbf{B}={J^4} \, , 
\end{eqnarray}
while the null eigenvector equations \eqref{eq:J_null_5d} {are the same as} \eqref{eq:null_split_0}--\eqref{eq:null_split_a}. {As in the free case, the Euler-Lagrange equations \eqref{phidot-B_J}--\eqref{EB_J} form an independent set, and the null eigenvector equations \eqref{eq:null_split_0}--\eqref{eq:null_split_a} are a consequence of that set. But again, other combinations can be chosen as independent, such as  our advection formulation, which consists of the set of five equations   \eqref{eq:null_split_4}, \eqref{eq:null_split_a}, and \eqref{B.Del phi_J}.
Explicitly, {for the sake of clarity of presentation,} the advection formulation reads}
\begin{eqnarray}
\label{eq:3d_red_constraint_J}
\mathcal{L}_{\mathbf{B}} A_4 &=& \rho\,,\\ 
\label{eq:3d_red_A4_J}
(\partial_0 + \mathcal{L}_{\bt{V}})A_4 &=& 0\,,\\
\label{eq:3d_red_advection_J}
(\partial_0 + \mathcal{L}_{\bt{V}})\bt{A} &=& \nabla (A_0 + V^b A_b) + V^4 \nabla A_4\,.
\end{eqnarray}
{We remark that the left-hand side of Eq. \eqref{EB_J} is the Pontrjagin density, which is found to coincide in our model with the fifth component of the conserved current $J^\mu$. This could have an application in models with non everywhere-vanishing Pontrjagin density, such as, for example, in terms of the electrodynamics of plasma surrounding pulsars \citep{goldreich1969pulsar}.}

{The conservation law \eqref{eq:J_5d_cons} is a necessary condition following from the above systems: namely, it follows from the independent system \eqref{phidot-B_J}--\eqref{EB_J} and it also follows from the independent system \eqref{eq:3d_red_constraint_J}--\eqref{eq:3d_red_advection_J}.} In the reduction to $3+1$ dimensions, if all the fields are invariant under translations in the direction $x^4$,  it becomes the continuity equation for the {external charge} density,
\begin{equation}
\label{rho_cons}
\partial_0 \rho + \nabla \cdot (\bt{V} \rho) = 0\,.
\end{equation}
{The proto-helicity, equation \eqref{eq:j_heli_def}, is not necessarily conserved in this formulation. In fact, the divergence of the proto-helicity is still given by equation \eqref{eq:j_heli_div}, but now we use equations \eqref{phidot-B_J}--\eqref{EB_J} to obtain
\begin{equation}
\label{eq:j_heli_div_ext_J}
  \partial_{\mu}{\heli}^{\mu} =  2\,J^4 =  2\,V^4 \rho\,.
\end{equation}
 We will discuss this in more detail in the next section, where we define a set of variables pertaining to a fluid, such as velocity, mass density, entropy, temperature and helicity.}

%%%%%%%%%%%%%%%%%%%%%%%%%%%%%%%%%%%%%%%%%%%%%%%%%%%
\section{{Promoting $J^\mu$ to a dynamical variable: {Fluid dynamics formulation}}} % 4 %
\label{sec:flu_dyn}
%%%%%%%%%%%%%%%%%%%%%%%%%%%%%%%%%%%%%%%%%%%%%%%%%%%
{\subsection{Reduction to $3+1$ dimensions}
When searching for a physical interpretation of a model, one usually relies on pattern matching. Our heuristic thought process goes as follows:
\begin{enumerate}
    \item Since our Chern-Simons model produces advection equations along the vector field $\bt{V} = (V^1, V^2, V^3)$, the argument in favour of interpreting this as the velocity field of a fluid is quite compelling. Thus, the first idea is to introduce a dynamical coupling between the connection components $(A_1, A_2, A_3)$ and the velocity field components $(V^1, V^2, V^3)$ {via an equation of the form $V^a = \ell g^{ab}(A_b + \partial_b \Phi)$, where $\ellfive$ is a constant with units of length, $\Phi$ is a scalar related to gauge invariance and $g$ is the (given) Euclidean metric,} so that the advection equation \eqref{eq:3d_red_advection_J} becomes the Euler fluid equation for the velocity field $\bt{V}$. 
    \item The second idea is that, because the fluid is not necessarily incompressible, we need to include thermodynamic variables (mass density, specific entropy, pressure, temperature) into the description. Notice that the advection equation \eqref{eq:3d_red_A4_J} refers to the conservation of the scalar $A_4$ along the pathlines of the fluid's velocity field. Amongst the thermodynamic variables, in the dissipationless case it is the specific entropy that is conserved in such a way. Thus, we are compelled to identify $A_4$ with the specific entropy $s$, via $s = \ellfive A_4$.   
    Next, the component $V^4$ of the velocity field in the fifth dimension is, in a sense, the ``conjugate variable'' to $A_4$, and the latter is proportional to the specific entropy. It will turn out that the thermodynamic relations will force us to interpret $V^4$ as the temperature $T$ of the fluid.
    \item The third idea is to find a physical interpretation for the charge density $\rho = J^0$. The fact that it satisfies \eqref{rho_cons} suggests that $\rho$ could be related to the (non-negative) mass density $\rho_{\mathrm{mass}}$. But how is this relation accomplished? The constraint \eqref{eq:3d_red_constraint_J}, written in terms of the velocity $\bt{V}$ and the specific entropy $s$, gives $\ellfive^2 \rho = (\nabla \times \bt{V}) \cdot \nabla s$, which looks like the definition of the Rossby-Ertel's potential vorticity.  Explicitly, one should factorise $\rho = \rho_{\mathrm{mass}} C/\ellfive^2$, leading to the well-known definition of potential vorticity charge: $C = (\nabla \times \bt{V}) \cdot \nabla s/ \rho_{\mathrm{mass}}$. Thus, given that $\bt{V}$ and $s$ satisfy the Euler equations of a dissipationless fluid, one concludes that $C$ is a conserved scalar charge if and only if $\rho_{\mathrm{mass}}$ satisfies the continuity equation, which further supports this interpretation: in particular, it is consistent with equation \eqref{rho_cons}.
    \item Finally, we need to implement points (i)--(iii) in the robust context of an action principle. We look for a minimalistic approach. The main problem we face here is that the factorization  $\rho = \rho_{\mathrm{mass}} C/\ellfive^2$ discussed in point (iii) is not ``natural'', as neither $\rho_{\mathrm{mass}}$ nor $C$ are fundamental fields: namely, the single scalar $\rho$ constitutes just one degree of freedom, while in principle $\rho_{\mathrm{mass}}$ and $C$ constitute $2$ degrees of freedom. So, the simplest approach is to choose one of the latter as dynamical, and the other one as secondary. Because the kinetic and internal energy densities of a fluid depend on $\rho_{\mathrm{mass}}$ and not on $C$, we choose $\rho_{\mathrm{mass}}$ to be the dynamical variable, and $C$ to be a derived quantity. Thus, we set $\rho_{\mathrm{mass}} := |\rho|$ and $C = \ellfive^2 \mathrm{sign}(\rho)$. Notice that in general it is expected that $\rho=0$ in spacetime regions of codimension $1$, which means that for a given time we expect $\rho=0$ in spatial surfaces. These surfaces will separate spatial regions of positive-definite and negative-definite $\rho$. Thus, when attempting to construct an action principle that depends on $\rho$ in a differentiable manner, we need to ensure that the thermodynamics is consistent: both kinetic energy and internal energy densities must be odd functions of $\rho$.       
\end{enumerate}}

{
\subsection{Constructing an action principle for the Chern-Simons fluid dynamics formulation}
\label{subsec:CS_fluid_action}
We wish to build up our action principle starting from the action given in  equation \eqref{int-action}.} 
{A feature of the system that one would like to preserve is the conservation law for $J^\mu$. This is easily achieved by including an additional scalar field $\Phi$ through the combination
$   J^\mu \partial_\mu \Phi $ {in the Lagrangian density,}
so that varying with respect to $\Phi$ enforces $\partial_\mu J^\mu =0$. {In addition, this scalar field $\Phi$ will ensure gauge invariance of the theory.}}
Next, the evolution of the dynamical field $A_\mu$ should be expected to affect the resulting current $J^\mu$ that couples to it. One way to ``upgrade" ${J}^\mu$ to a dynamical variable is to add a term $\mathcal{U}(J^\mu)$ to the {original Lagrangian density in \eqref{int-action}, where $\mathcal{U}$ is a differentiable function,} so that varying with respect to $J^\mu$ yields meaningful equations that link the source to the dynamical fields.

{The dynamical equations for $J^\mu$ are then determined by $\mathcal{U}(J^\mu)$. This term necessarily requires a metric structure. Some natural choices would be Minkowski space-time and non-relativistic Euclidean space, among others. In the latter case, for example, since the spatial components of $J$ correspond to a momentum density, one can define  
\begin{equation}
\label{eq:kinetic+potential}
\mathcal{U}(J^\mu) = \frac{1}{2\ellfive}\frac {J_a J^a}{J^0}- \frac 1 \ellfive \mathcal{E}(J^0,J^4)\,,
\end{equation}
where $a$  {is summed over from $1$ to $3$} and $\ellfive$ is the length of the circle parameterised  by the fifth coordinate, $x^4$. Here, the term quadratic in momentum corresponds to the kinetic energy {(note how its dependence on $J^0$ reveals the importance of spacetime regions where $J^0 \neq 0$, as mentioned earlier)}, while the function $\mathcal{E}$ that relies on the remaining components $J^0\,, J^4$, represents the internal energy density of the fluid. This choice breaks space-time covariance and is justified in a non-relativistic setting.  {Notice that our choice \eqref{eq:kinetic+potential} basically sets $\rho (= J^0)$ to represent the mass density, up to a sign, because the kinetic energy density reads $\frac{1}{2\ellfive}\frac {J_a J^a}{J^0} = \frac{\rho}{2\ellfive} |\bt{V}|^2$. Thus, we define the mass density as 
\begin{equation}
    \label{eq:mass_density}
    \rho_{\mathrm{mass}} := |\rho|\,,
\end{equation} 
which gives the factorization $\rho = \mathrm{sign}(\rho)\rho_{\mathrm{mass}}$, that will set the potential vorticity charge to be uniform up to a sign. Also, because the kinetic energy is odd in $\rho$, in order to get a consistent thermodynamics we need to impose $\mathcal{E}(J^0,J^4)$ to be odd in $\rho$: 
\begin{equation}
    \label{eq:E_odd}
    \mathcal{E}(-J^0,-J^4) = -\mathcal{E}(J^0,J^4)\,,\qquad \text{for \quad any \quad} J^0, J^4 \in \mathbb{R}\,.
\end{equation}
These choices fulfil our heuristic points (iii)--(iv). }

Putting these assumptions together, one arrives at the following action functional\footnote{{Here, we can easily deduce the physical dimension, in natural units, of the quantities in the Action. It is as follows:
$\left[A_\mu\right]= L^{-1}\,,\;\left[J^\mu\right]= L^{-4}\,,\;\left[\Phi\right]= L^{0}\,,\;\left[\ellfive\right]= L^{1}\,,\;\left[\mathcal{E}\right]= L^{-4}$.
}} 
\begin{equation} \label{fullaction}
{I[A_\mu,J^\mu,\Phi] = \int_{{\mathcal{M}}_5} \left(\frac{1}{3!} A\wedge \mathrm{F}\wedge \mathrm{F} - \left[(A_\mu + \partial_\mu \Phi) J^\mu - \frac{1}{2\ellfive} \frac{J_a J^a}{J^0} + \frac {1}\ellfive\,\mathcal{E}(J^0,J^4) \right]\mathrm{d}^5x \right).}
\end{equation}}

{The Euler-Lagrange equations obtained varying with respect to $A, J^0, J^a,J^4$ and $\Phi$ read:
\begin{eqnarray}
\label{A}
 {\frac{1}{2}} \,\mathrm{F}\wedge \mathrm{F} &=& {-} {}^*{J}\,, \\
\label{J0}
A_0 + \partial_0 \Phi  &=& -\frac{J_a J^a}{2\ellfive(J^0)^2}  - \frac{1}{\ellfive} \frac{\partial}{\partial J^0} \mathcal{E}(J^0,J^4)\,,\\
\label{Ja}
A_a + \partial_a \Phi &=&\frac {J_a}{\ellfive J^0}  \,,\quad a = 1, 2, 3\,,\\
\label{J4}
A_4+ \partial_4 \Phi  &=&- \frac 1\ellfive\frac{\partial}{\partial J^4} \mathcal{E}(J^0,J^4)  \,,\\
\label{phi}
\partial_\mu J^\mu &=& 0\,.
\end{eqnarray}
Here we are allowing $\Phi$ and $A$ to depend on $x^4$ and we expect $|J^\mu/J^0|$ to remain bounded.}

{The kinetic term induces a direct relation between $J^a/J^0$ and the connection $A_a$ {via equations \eqref{Ja}, thus fulfilling our heuristic point (i).} A sensible relation between $J^\mu$ and $A_\mu$, however, should take into account the fact that $A_\mu$ is a gauge potential while $J^\mu$ is a gauge-invariant vector field. This mismatch is reconciled in \eqref{J0}--\eqref{J4} by the presence of $\Phi$, which is a St\"uckelberg field that combines the gauge field $A_\mu$ into gauge invariant expressions. In other words, $\Phi$ plays a double r\^ole: it enforces current conservation and allows for $J$ to couple to a {gauge-invariant ``dressed field" \citep{franccois2024dressing},} guaranteeing the preservation of gauge symmetry in the action.}

{The fact that $J^{\mu}=(J^0,J^a,J^4)$ is a null eigenvector of $\mathrm{F}$, which follows from equation \eqref{A}, leads to the following equations:
\begin{equation}
\label{energy}
J^a(\partial_a A_0 - \partial_0 A_a)  +J^4 (\partial_4 A_0 - \partial_0 A_4) = 0\,,
\end{equation}
\begin{equation}
\label{momentum}
\partial_0 A_a +\frac{1}{J^0} J^b \partial_b A_a = \partial_a A_0 + \frac{1}{J^0}J^b \partial_a A_b + \frac{1}{J^0} J^4 (\partial_a A_4 - \partial_4 A_a)\,,
\end{equation}
\begin{equation}
\label{cons_A4}
\partial_0 A_4 + \frac{1}{J^0} J^b(\partial_b A_4 - \partial_4 A_b) = \partial_4 A_0\,.
\end{equation}}

We recall that the last 4 of these 5 equations are independent. Thus, we will leave equation \eqref{energy} out of the following discussion. The time component of equations \eqref{A} remains independent of these 4 equations: {it is actually the constraint \eqref{eq:3d_red_constraint_J}, and will play an important role, deserving of the separate  subsection \ref{sec:Ertel}.
}

%%%%%%%%%%%%%%%%%%%%%%%%%%%%%%%%%%%%%%%%%%%%%%%%%%%%%%%%%%%%%
\subsection{{Euler fluid equations and thermodynamics}}
\label{sec:interacting}
%%%%%%%%%%%%%%%%%%%%%%%%%%%%%%%%%%%%%%%%%%%%%%%%%%%%%%%%%%%%
{While setting up the action principle in section \ref{subsec:CS_fluid_action} we have identified a couple of physical quantities pertaining to a fluid: the velocity field of components ${V}^a = J^a/J^0 $ and the mass density $\rho_{\mathrm{mass}} = |J^0|$. We are yet to fulfil our only remaining heuristic point (ii), namely to define the specific entropy $s$ and the temperature $T$ of the fluid. }
{So far, variables $A_4$ and $J^4$, that arise due to the fifth dimension, do not have an obvious interpretation {in terms of physical quantities in the three-dimensional space}. As will be shown, it is quite natural to interpret them as thermodynamic variables and to relate $\mathcal{E}$ to a thermodynamic potential. This description assumes that our fields be understood as averages{, } over microscopic counterparts{, } describing the physical system. These averages entail a scale which separates the macroscopic description we are dealing with from a microscopic one. In order to provide a thermodynamic interpretation one can imagine replacing the local fields $A_\mu(x), J^\mu(x), \Phi(x), \mathcal{E}(x)$ by their averages over a sufficiently large spacetime {neighbourhood around $x$, containing large numbers of {microscopic} particles to have a meaningful notion of temperature, density, entropy, and pressure.  These neighbourhoods} must also be small enough to be considered {as} points from a large scale viewpoint. See, e.g., \cite[Ch.3]{huang2008statistical}.}

{Substituting the expressions for $A_0,\,A_a\,, A_4$ from equations (\ref{J0})--(\ref{J4}) into (\ref{energy})--(\ref{cons_A4}) leads to 
\begin{equation}
  \left( \partial_0 + \mathbf{V}\cdot \mathbf{\nabla}\right) V_a=-\partial_a\left( \frac{\partial \mathcal{E}}{\partial J^0}\right) -\frac{J^4}{J^0}\partial_a \left(\frac{\partial \mathcal{E}}{\partial J^4}\right) {-\frac{J^4}{J^0}\partial_4 V_a}\,,
  \label{evol}
\end{equation}} 
{
\begin{equation}
(\partial_0 + \mathbf{V}\cdot \mathbf{\nabla})\left(\frac{\partial \mathcal{E}}{\partial J^4}\right)= \partial_4 \left(\frac{\partial \mathcal{E}}{\partial J^0}\right)\,.\label{entropyflow}
\end{equation}
}

{Notice that the dependence of the fields on the $x^4$-coordinate{, which could be problematic for the interpretation of the model in terms of Euler fluids in three-dimensional space,} will manifest itself only via the last terms of the above equations. Two potentially useful strategies are possible: (i) simply to assume that the fields do not depend on $x^4$, or (ii) to assume that the fields are periodic functions of $x^4$, with the same period (i.e., wavelength) $L < \infty$. The latter strategy is more complicated but it would allow us to take averages (over the $x^4$-period) of the above equations, leading to effective equations in the usual $(3+1)$-dimensional spacetime {which could model dissipation as flow along the fifth coordinate (see section \ref{sec:conclusions}).} Focusing on the first strategy {($\partial_4 :=0 $)} from here on, we see that equation  \eqref{entropyflow} states that the quantity $\frac{\partial \mathcal{E}}{\partial J^4}$ evolves in time but is conserved along the flow pathlines, {suggesting a relation with the specific entropy.}}
{Next, expression \eqref{evol} looks formally like the Euler equation for the velocity of an inviscid compressible fluid,
\begin{equation}\label{Inviscid}
\partial_0 \bt{V} +  (\bt{V} \cdot \nabla) \bt{V}  = - \frac{1}{\rho} \nabla p \,,
\end{equation}
{in terms of the charge density $J^0=\rho$ and the quantity $p$ related to the fluid pressure}, with 
\begin{equation}
\nabla p := \rho \,\nabla\left(\frac{\partial \mathcal{E}}{\partial \rho}\right) + J^4\,\nabla \left(\frac{\partial \mathcal{E}}{\partial J^4}\right).
\end{equation}
Now, since $\mathcal{E}=\mathcal{E}(\rho, J^4)$, this last relation can be written as
\begin{equation}\label{nabla p}
\nabla p = \nabla\left(\rho\,\frac{\partial \mathcal{E}}{\partial \rho} + J^4\,\frac{\partial \mathcal{E}}{\partial J^4} -\mathcal{E} \right) \,,
\end{equation}
from which it follows that, up to a constant, 
\begin{equation} \label{p}
p = \rho\,\frac{\partial \mathcal{E}}{\partial \rho} + J^4\,\frac{\partial \mathcal{E}}{\partial J^4} -\mathcal{E} \,.
\end{equation}
{Notice that our condition that $\mathcal{E}$ be an odd function of $\rho$, equation \eqref{eq:E_odd}, is indeed a consistency requirement, as it implies that the function $p$ defined above is an odd function of $\rho$, and thus our definition of mass density \eqref{eq:mass_density} implies that the Euler fluid equation \eqref{Inviscid} is valid in regions of positive $\rho$ as well as in regions of negative $\rho$: the fluid behaves as a normal fluid in both types of region, with mass density $\rho_{\mathrm{mass}} = |\rho| = \mathrm{sign}(\rho) \rho$ and pressure $p(\rho_{\mathrm{mass}},\,T) = \mathrm{sign}(\rho) p(\rho,\, T)$, so that the right hand side of equation \eqref{Inviscid} is simply a function of $\rho_{\mathrm{mass}}$, and no negative-pressure paradox occurs. Having understood this point, } 
{we now discuss the thermodynamic interpretation rendered by expression \eqref{p}. Let us assume an equilibrium condition described by the thermodynamic variables $\rho,T$, corresponding to the three-dimensional variables mass density (or charge density, depending on the interpretation) and temperature, respectively. 
Hence, one would interpret the energy density $\mathcal{E}$ in terms of the thermodynamic potential $f$, function of the thermodynamic variables:
\begin{equation}\label{ef}
\mathcal{E}(\rho,J^4(\rho,T)) := \rho f(\rho,T)  \,,
\end{equation}
in which one identifies $f(\rho,T)$ as the specific Helmholtz free energy (namely, the Helmholtz free energy per unit mass or per unit charge, depending on the interpretation).}
Now,
\begin{align}
\mathrm{d}\mathcal{E}&= \frac{\partial \mathcal{E}}{\partial \rho}\mathrm{d}\rho + \frac{\partial \mathcal{E}}{\partial J^4}\mathrm{d}J^4 = \left(\frac{\partial \mathcal{E}}{\partial \rho} + \frac{\partial \mathcal{E}}{\partial J^4}\frac{\partial J^4}{\partial \rho}\right) \mathrm{d}\rho + \frac{\partial \mathcal{E}}{\partial J^4}\frac{\partial J^4}{\partial T} \mathrm{d}T\,,
\end{align}
{and the first law of thermodynamics, in the form of the %well-known 
relation $\mathrm{d}f = - p \mathrm{d}(1/\rho) - s \mathrm{d}T$} applied to $f=\mathcal{E}/\rho$ gives, for the pressure $p$ and the specific entropy $s$,
\begin{align}
\label{-p}
p := \rho^2 \frac{\partial f}{\partial \rho} & = -{\mathcal{E}}  + {\rho} \left(\frac{\partial \mathcal{E}}{\partial \rho} + \frac{\partial \mathcal{E}}{\partial J^4}\frac{\partial J^4}{\partial \rho}\right) \,,\\ 
\label{-S}
s := - \frac{\partial f}{\partial T} & = - \frac{1}{\rho}  \frac{\partial \mathcal{E}}{\partial J^4} \frac{\partial J^4}{\partial T} \,.
\end{align}

Now, comparing the previous relation \eqref{p} with the relation \eqref{-p}{, using the expression \eqref{ef},} we get, after simplifications,
$$  \left(J^4-{\rho}  \frac{\partial J^4}{\partial \rho}\right)\,\frac{\partial \mathcal{E}}{\partial J^4}  =  0\,,$$
which is a differential equation for $J^4$: assuming $\frac{\partial \mathcal{E}}{\partial J^4} \neq 0$ we get
 $J^4 = \rho I(T)\,,$ where $I$ is some function to be determined. As for equation \eqref{-S}, it becomes, after replacing the above formula for $J^4$,
$$ s= -\frac{\partial \mathcal{E}}{\partial J^4} I'(T)\,,$$
and this suggests that we set $I(T) = T$, where we have set to unity the proportionality constant. In summary, our thermodynamic interpretation
\begin{align} 
\label{J0_summary}
J^0 &= \rho\,,\\
\label{J4_summary}
J^4 &= \rho T \,,\\
\label{E_summary}
\mathcal{E}(\rho,J^4) &= \rho f(\rho,T)\,,
\end{align}
simplifies equations \eqref{-p}--\eqref{-S} to
\begin{align} \label{-S/N}
\frac{\partial \mathcal{E}}{\partial J^4} &=- s \,,\\
\label{H/N}
\frac{\partial \mathcal{E}}{\partial \rho} &= \frac{p}{\rho} + f + T s = e + \frac{p}{\rho} = h \,,
\end{align}
where $e := f + T s$ is the specific internal energy and $h := e + \frac{p}{\rho}$ is the specific enthalpy. 
\\
With these identifications, we can go back to eqs. \eqref{evol}, \eqref{entropyflow} and express them as:
\begin{align} \label{V2}
\partial_0 \bt{V} +  (\bt{V} \cdot \nabla) \bt{V}  &= - \frac{1}{\rho} \nabla p = -\nabla h + T \nabla s
\,,\\ 
\label{SH/N}
\left(\partial_0 +\mathbf{V}\cdot \mathbf{\nabla}\right) s  &=0\,
\end{align}
which, when considered along with the continuity equation \eqref{rho_cons} (one of our Euler-Lagrange equations \eqref{phi}), constitute the standard fluid equations for a system without viscosity, heat sources and external forces. {Of course, we must remark that $\rho$, being a charge density, can take positive and negative values, but from the fact that $\mathcal{E}$ is an odd function of $\rho$ it follows that $e, f, h$ and $s$ are even functions of $\rho$, while $p$ is an odd function of $\rho$, resulting in the usual thermodynamics if we now work with the mass density $\rho_{\mathrm{mass}} = |\rho|$.}} 

{To complete the above picture, there is only one remaining equation to consider:} the constraint $\frac{1}{8}\epsilon^{jklm}\mathrm{F}_{jk}\mathrm{F}_{lm}= \rho$, namely the $0$-component of equation \eqref{A}.

\subsection{Understanding the constraint $\frac{1}{8}\epsilon^{jklm}\mathrm{F}_{jk}\mathrm{F}_{lm}= \rho$}

\subsubsection{The constraint $\frac{1}{8}\epsilon^{jklm}\mathrm{F}_{jk}\mathrm{F}_{lm}= \rho$ and the Rossby-Ertel's invariant (a.k.a. potential vorticity charge)}
\label{sec:Ertel}

In terms of the physical variables, the constraint reads
\begin{equation}
\label{eq:constraint}
    (\nabla \times \bt{V}) \cdot \nabla s = \ellfive^2 \rho\,,
\end{equation}
{which determines a class of flows: because of our choice $\rho_{\mathrm{mass}} = |\rho|$ we get $\rho = \rho_{\mathrm{mass}} \mathrm{sign}(\rho)$, so the constraint becomes
\begin{equation}
\label{eq:constraint_phys}
\frac{(\nabla \times \bt{V}) \cdot \nabla s}{ \rho_{\mathrm{mass}}} = \ellfive^2 \mathrm{sign}(\rho)\,,
\end{equation}
but the left hand side of this equation is the Rossby-Ertel's potential vorticity charge, a known invariant \citep{haynes1990conservation,kurgansky2002adiabatic,mcintyre2003potential}, and thus we obtain a class of flows defined by the particular fact that the potential vorticity charge must be uniform throughout the fluid, up to changes in sign, but with constant magnitude, given by the square of the length of the fifth dimension, a very curious result. The constraint, written in the ``physical'' form (\ref{eq:constraint_phys}), appears as a simple condition on the initial data, which nevertheless imposes strong restrictions on the type of flows admitted by our model: in regions of significant mass density $\rho_{\mathrm{mass}}$, there must be significant vorticity $\nabla \times \bt{V}$ and significant gradients of specific entropy $\nabla s$, so that the ratio in \eqref{eq:constraint_phys} remains uniform.} 

Notice that, as is well known, the conservation of the Rossby-Ertel's potential vorticity charge $C := \frac{1}{\rho_{\mathrm{mass}}} \,\mbf{\omega} \cdot \nabla s$ (where $\mbf{\omega} = \nabla \times \bt{V}$ is the vorticity field) over time can be derived from the evolution equations (\ref{rho_cons}), (\ref{V2}), and (\ref{SH/N}): it follows from the facts that $s$ is conserved and that $\frac{1}{\rho_{\mathrm{mass}}} \,\mbf{\omega}$ is a vector field (not a density) satisfying
$$\left(\partial_0 + {\mathcal L}_{\bt{V}} \right) \left(\frac{1}{\rho_{\mathrm{mass}}} \mbf{\omega}\right) = - \frac{1}{\rho_{\mathrm{mass}}}\nabla s \times \nabla T\,.$$

\subsubsection{The constraint $\frac{1}{8}\epsilon^{jklm}\mathrm{F}_{jk}\mathrm{F}_{lm}= \rho$ and regions of sign-definite charge density $\rho$}
\label{sec:sign-definite}
{Let us discuss in more detail the justification for our interpretation of  $\rho_{\mathrm{mass}} = |\rho|$ as the mass density of a fluid. 
Consistency of \eqref{eq:constraint} implies that $\rho$ cannot be interpreted as the scalar mass density since the vorticity field is an axial vector. This problem was also encountered in the Chern-Simons atmospheric model by \cite{jacques2022simplified}. In that case, the right hand side of \eqref{eq:constraint} was defined as $\beta j^0$, where $\beta$ stands for an odd function of the latitude ($\sim \sin \theta$) that compensates for the sign change of the Coriolis force in different hemispheres and $j^0$ is the bona fide density of the fluid. Further, notice that we can write \eqref{eq:constraint} as
\begin{equation}
\label{eq:monopole}
\nabla \cdot \mbf{\widetilde{B}} = \rho \,,\qquad \mbf{\widetilde{B}} :=\ellfive^{-2} s \nabla \times \bt{V}\,,
\end{equation}
where $\mbf{\widetilde{B}}$ would play the role of a ``magnetic field'' if $\rho$ were the density of magnetic monopoles. In this case there would be no conflict with $\rho$ being a pseudo scalar.}

{Because of the general form \eqref{eq:monopole} of the constraint as a divergence, at a given time $t$ the spatial manifold $\mathcal{M}_3(t)$ is separated into regions of positive values of $\rho$  and regions of negative values of $\rho$:} regions with sign-definite $\rho$ are advected by the flow, because from equation (\ref{rho_cons}) it follows that the zeroes of this charge density are ``conserved'' and advected by the flow. Moreover, the zeroes of the density are generically, at any given time, $2$-dimensional surfaces embedded in $\mathcal{M}_3$: the equation $\rho(x,y,z,t)=0$ defines, at a given time $t$, these surfaces, which are then advected by the flow over time, and separate the regions of positive-definite charge density from the regions of negative-definite charge density. 

Now, recalling the discussion in section \ref{sec:J^mu} about the degeneracy of the spacetime points where $(J^0 =)\rho=0$, namely that there are no dynamical degrees of freedom propagating in such places, we conclude that for all purposes, the surfaces $\rho=0$ separate the spacetime into independent dynamical regions with sign-definite charge density. {As we will see in section \ref{sec:cons}, these regions do not exchange matter or energy, but can exchange helicity.}

{As for the consistency of our model's fluid equations and thermodynamics,} in terms of the equations of motion for the main physical variables $\rho, s, \bt{V}, T$, we recall that the simple extension that $f$ be an even function of $\rho$ (equivalently, that $\mathcal{E}$ be an odd function of $J^\mu$ in the action principle) leads to the usual fluid equations on each region of sign-definite $\rho$, with ``physical'' mass density $\rho_{\mathrm{mass}}=|\rho|$, pressure $p(|\rho|,T)$ and specific entropy $s(|\rho|,T)$. Namely, within a given region of sign-definite density $\rho$, there is no way to tell whether $\rho$ is positive or negative by looking at the dynamics. {Only when looking at the constraint \eqref{eq:constraint}, it is possible to tell the difference, because the Rossby-Ertel's potential vorticity charge will take different signs. However, we remark that,} as explained in \citep{kurgansky2002adiabatic} for an atmospheric model involving southern and northern hemispheres, even the overall sign of the potential vorticity charge (and thus the sign of our $\rho$) is a matter of convention, as it depends on the choice of chirality of the velocity field.

{\subsection{Connection with 
Crocco's equation and Kuznetsov's Lorentz-force hydrodynamics}
\label{sec:Crocco}
For this analysis we restore the dependence on the $x^4$ variable. Equations \eqref{J0},\eqref{Ja},\eqref{J4} give:}
\begin{eqnarray}
     A_0 +\partial_0 \Phi &=& -\frac 1{\ellfive}\left(\frac 12 \mathbf{V}^2 +h
    \right)\,,\\
     \mathbf{A} +\mathbf{\nabla} \Phi &=& \frac 1{\ellfive}\mathbf{V}\,,\\
     A_4 +\partial_4\Phi &=& \frac 1{\ellfive}s\,.
\end{eqnarray}
Up to a factor, the right-hand sides of these  equations correspond respectively to  {the Bernoulli function \citep{vallis2017atmospheric} (a.k.a. the stagnation enthalpy \citep{currie2002fundamental}), the velocity of the fluid, and the specific entropy.}

{The corresponding field-strengths, on-shell turn out to be:
\begin{align}
\label{eq:E}
\mathbf{E}& := -\mathrm{F}_{0a}\hat{\mathbf{x}}^a=- \frac 1\ellfive\left[
\partial_0 \mathbf{V}+\mathbf{\nabla}\left(\frac 12 \mathbf{V}^2 +h
\right)\right]
 \\
\label{eq:B}
\mathbf{B}& := \frac 12\epsilon_{abc}\mathrm{F}_{bc}\hat{\mathbf{x}}^a= \frac 1\ellfive \mathbf{\nabla}\times \mathbf{V}
 \\
\label{eq:Fa4}
 \mathrm{F}_{a4}\hat{\mathbf{x}}^a&= \frac 1\ellfive\left[\mathbf{\nabla}s-\partial_4\mathbf{V}\right]
 \\
\label{eq:F04}
\mathrm{F}_{04}&=\frac 1\ellfive \left[
\partial_0 s +{\partial_4\left(\frac 12 \mathbf{V}^2 +
h\right)}
\right]\,.
\end{align}}
{We recall here that our fluid-dynamical equations (\ref{V2}) and (\ref{SH/N}) are simply a rewrite of the fact that the current $J^\mu$ is a null vector of the field-strength $\mathrm{F}_{\mu \nu}$. Thus, it is instructive to look at these equations once again from the perspective of the field-strength components. First, the momentum conservation equations (\ref{V2}) are actually
$$-\frac{\ellfive}{J^0} \mathrm{F}_{a \nu} J^\nu = 0\,, \quad a=1,2,3,$$
which we can decompose further as:
\begin{equation}
\label{eq:Crocco}
    -\ellfive(\mathrm{F}_{a 0} + \mathrm{F}_{a b} V^b + \mathrm{F}_{a 4} V^4) = 0\,, \quad a=1,2,3,
\end{equation}
and in terms of the formulae \eqref{eq:E}--\eqref{eq:Fa4} 
this gives, directly,
$$(\partial_0 + T \partial_4) \mathbf{V}+\mathbf{\nabla}\left(\frac 12 \mathbf{V}^2 +h
\right) - \mathbf{V} \times (\mathbf{\nabla}\times \mathbf{V}) - T \nabla s  = 0\,,$$
which is (in the $x^4$-independent case) Crocco's equation \cite[Ch.3]{currie2002fundamental}, \citep[Ch.8]{schaum_fluids_1999}.}

{Moreover, if we go back to equation \eqref{eq:Crocco}, we see that the combinations 
$\mathrm{F}_{a 0} + \mathrm{F}_{a b} V^b \,, \,\, a=1,2,3,$
are the components of $\mathbf{E} + \mathbf{V}\times \mathbf{B}$, namely the Lorentz force per unit charge. Thus, using equations \eqref{eq:E}--\eqref{eq:Fa4} and rearranging we get
$$(\partial_0 + T \partial_4) \bt{V} +  (\bt{V} \cdot \nabla) \bt{V} = - \ellfive (\mathbf{E} + \mathbf{V}\times \mathbf{B}) - \nabla h\,,$$
which is (in the $x^4$-independent case) the statement that the flow has negative {``electric'' charge, of magnitude $\ellfive$ per unit mass,} and moves under the influence of a Lorentz-force law combined with a conservative force where the specific enthalpy $h$ plays the role of potential energy per unit mass. We remark that this interpretation of the Euler fluid equations in terms of a Lorentz force resembles an earlier interpretation provided by \cite{kuznetsov2008mixed}, where, {at difference with our approach,} the velocity field is decomposed in terms parallel plus normal to the vorticity field.}

{\subsection{Conservation laws of the fluid dynamics formulation of the Chern-Simons model} 
\label{sec:cons}
The fluid dynamics formulation of our Chern-Simons model, in the case when the fields do not depend on the fifth coordinate $x^4$, is identical to the dissipationless $(3+1)$-dimensional Euler equations for a compressible fluid with general thermodynamics. Thus, it features the usual conservation laws of fluids: mass, energy, entropy, potential vorticity, and helicity.\\
Now, because of the constraint \eqref{eq:constraint}, the potential vorticity charge density $\rho$ is written as a divergence of the specific entropy times the vorticity, and this will have several implications. In addition, helicity is not conserved for baroclinic fluids but its production term and flux terms get simplified due to the constraint.\\
Some of the results of this subsection (which we will clearly note) apply to any dissipationless compressible $(3+1)$-dimensional fluid. All other results in this subsection apply to any dissipationless compressible $(3+1)$-dimensional fluid such that the constraint \eqref{eq:constraint} is satisfied initially.\\ 
In the remainder of this subsection, we will use the Leibniz-Reynolds transport theorem, which states
\begin{equation}
\label{eq:Leibniz}
    \frac{\mathrm{d}}{\mathrm{d} t} \left(\int_{m_3(t)}U(\mathbf{x},t)\mathrm{d}^3x\right) = \int_{m_3(t)}\left[\partial_0 U(\mathbf{x},t) + \nabla\cdot(\mathbf{V}U(\mathbf{x},t))\right]\mathrm{d}^3x\,,
\end{equation}
for any sufficiently smooth scalar density field $U(\mathbf{x},t)$, where $m_3(t) (\subset \mathcal{M}_3(t) \subset \mathbb{R}^3)$ is a generic open region that is advected by the velocity field $\bt{V}$.

}

\subsubsection{Energy conservation in regions of sign-definite potential vorticity charge density $\rho$} 

{The results in this subsubsection apply to any dissipationless compressible $(3+1)$-dimensional fluid, irrespective of the constraint \eqref{eq:constraint}.} The energy equation is a consequence of the continuity equation (\ref{rho_cons}), the momentum equations (\ref{Inviscid}) and the first law of thermodynamics $\mathrm{d}e = \frac{p}{\rho^2}\mathrm{d}\rho + T \mathrm{d}s$, where $e = f+Ts$ is the specific internal energy. It reads (we omit the proof for simplicity, as it is standard):
\begin{equation}
\label{eq:energy_cons}
\partial_0 \left(\frac{\rho |\bt{V}|^2}{2}  + \rho \, e\right) + \nabla \cdot \left(  \bt{V} \left[ \frac{\rho |\bt{V}|^2}{2}  + \rho \, e +  p\right]   \right) = 0\,.
\end{equation}
Notice that in our extension to regions where  $\rho<0$, the fact that $f$ is an even function of $\rho$ implies that $e$ is an even function of $\rho$. Thus, the above energy conservation equation is satisfied by replacing $\rho \to |\rho|$ in each sign-definite region. It is instructive, though, to focus on just a single sign-definite open region $m_3^\rho(t) \subset \mathbb{R}^3$, such that $\rho(\mathbf{x},t)$ has a definite sign for $\mathbf{x}\in m_3^\rho(t)$ and such that $\rho(\mathbf{x},t) = 0$ for $\mathbf{x}\in \partial m_3^\rho(t)$. As explained in section \ref{sec:sign-definite}, {such a region} evolves in time, being advected by the flow $\mathbf{V}$, see also Appendix \ref{lastappendix}. Thus, it is appropriate to apply the Leibniz-Reynolds transport theorem \eqref{eq:Leibniz} to the scalar density $U = \frac{\rho |\bt{V}|^2}{2}  + \rho \, e$ and using equation (\ref{eq:energy_cons}) we get
$$\frac{\mathrm{d}}{\mathrm{d} t} \left(\int_{m_3^\rho(t)}\left(\frac{\rho |\bt{V}|^2}{2}  + \rho \, e\right)\mathrm{d}^3x\right) = {-}\int_{m_3^\rho(t)}\nabla\cdot(\mathbf{V}p)\mathrm{d}^3x = {-}\oint_{\partial m_3^\rho(t)}p \mathbf{V}\cdot \mathbf{\hat{n}}\mathrm{d}^2x\,,$$
which is equal to zero since $p=0$ when $\rho=0$ by virtue of it being an odd function of $\rho$. Thus we conclude that the total energy in each region $m_3^\rho(t)$ is conserved independently.

{
\subsubsection{Mass conservation in regions of sign-definite potential vorticity charge density $\rho$} 
\label{sec:mass_cons}
The conservation law \eqref{rho_cons}, along with our definition of mass density $\rho_{\mathrm{mass}} = |\rho|$, implies the usual mass conservation law
$\partial_0 \rho_{\mathrm{mass}}  + \nabla \cdot \left(  \bt{V} \rho_{\mathrm{mass}}   \right) =0$, from which all known results of mass conservation follow, such as the constancy of the total mass $\int_{m_3(t)}\rho_{\mathrm{mass}}\mathrm{d}^3x$ inside a region $m_3(t)\subset \mathbb{R}^3$ that is advected by the fluid (without imposing any boundary condition). This result applies to any dissipationless compressible $(3+1)$-dimensional fluid, irrespective of the constraint \eqref{eq:constraint}.\\
Now, the presence of the constraint \eqref{eq:constraint} allows us to consider several scenarios that were not possible in the general case:\\
\textbf{Scenario 1.} A closed isentropic surface $s=s_0$ (constant). Suppose there is a closed surface $\Sigma(t)$ at time $t$, such that $s=s_0$ (constant) on $\Sigma(t)$. This would be true, for example, near local minima or maxima of $s$. Now, because $s$ is conserved, we know that $m_3^s(t) := \mathrm{int}(\Sigma(t))$ is advected by the velocity field $\bt{V}$. Thus, the Leibniz-Reynolds theorem  \eqref{eq:Leibniz} applies for the density $U:= \rho$, and thus the corresponding ``charge'' $M_0:= \int_{m_3^s(t)}\rho(\mathbf{x},t)\mathrm{d}^3x$ is conserved. Moreover, the constraint \eqref{eq:constraint} gives, in terms of the vorticity field $\mbf{\omega} = \nabla \times \bt{V}$, 
$$\ellfive^{2} M_0 = \int_{m_3^s(t)}\nabla \cdot (s \mbf{\omega})\mathrm{d}^3x = \oint_{\Sigma(t)}s \mbf{\omega}\cdot \hat{n} \mathrm{d}^2x = s_0 \oint_{\Sigma(t)}\mbf{\omega}\cdot \hat{n} \mathrm{d}^2x = 0\,.$$
But we have split $\rho = \rho_{\mathrm{mass}} \mathrm{sign}(\rho)$ so that $\int_{m_3^s(t)}\rho_{\mathrm{mass}} \mathrm{sign}(\rho)\mathrm{d}^3x = 0$, namely, whenever the mass density $\rho_{\mathrm{mass}}$ is not identically zero in $m_3^s(t)$, there must be two pieces containing mass: $m_3^{\pm}(t) := \{\bt{x} \in m_3^s(t) | \pm \rho(\bt{x},t) > 0\}$, with respective masses 
$M_\pm(t) := \int_{m_3^\pm(t)}\rho_{\mathrm{mass}} \mathrm{d}^3x$ which must be equal: $M_+(t) = M_-(t)$. Geometrically, this implies that the surface $\rho=0$ must intersect $m_3^s(t)$, separating it into two disjoint non-empty pieces $m_3^\pm(t)$. See figure \ref{fig:isentropic}.\\
Moreover, we notice that $M_\pm(t)$ are actually conserved: $\mathrm{d}M_\pm(t)/\mathrm{d}t = 0$.
This follows from the property of the boundary of $m^\pm_3$ of being the union of a surface of constant entropy and a surface at $\rho=0$, so that, as proven in  Appendix \ref{lastappendix}, these volumes are separately advected by the fluid velocity ${\bf V}$. 
Let us study in more detail the formula for this mass. Using figure \ref{fig:isentropic} as a reference and defining the constant $M:= M_+(t) = M_-(t)$ we can write
$$M = \int_{m_3^+(t)}\rho \mathrm{d}^3x = \oint_{\partial m_3^+(t)}s\mbf{\omega}\cdot \hat{n} \mathrm{d}^2x = \int_{\Sigma^+(t)}s\mbf{\omega}\cdot \hat{n} \mathrm{d}^2x + \int_{\Sigma^0(t)}s\mbf{\omega}\cdot \hat{n} \mathrm{d}^2x \,,$$
where $\Sigma^+(t) := \Sigma(t) \cap m_3^+(t)$ and $\Sigma^0(t) := \{\bt{x} \in m_3^s(t) | \rho(\bt{x},t) = 0\}$. Now, because $s=s_0$ on $\Sigma^+(t)$ we can rewrite the first term of the RHS in the above expression for $M$ as $s_0\int_{\Sigma^+(t)}\mbf{\omega}\cdot \hat{n} \mathrm{d}^2x = -s_0 \int_{\Sigma^0(t)}\mbf{\omega}\cdot \hat{n} \mathrm{d}^2x$, which gives finally
$$M = \int_{\Sigma^0(t)}(s-s_0)\mbf{\omega}\cdot \hat{n} \mathrm{d}^2x\,.$$
In summary, for a given value of entropy $s_0$ achievable by the fluid, one defines regions $m_3^\pm(t)$ (with opposite potential vorticity charges) which are advected by the fluid and such that their masses are constant, equal to each other and given by a $2$-dimensional boundary integral along the zero-mass surface $\rho_{\mathrm{mass}} = 0$.\\
\begin{figure}
    \centering
    \includegraphics[width=0.7\linewidth]{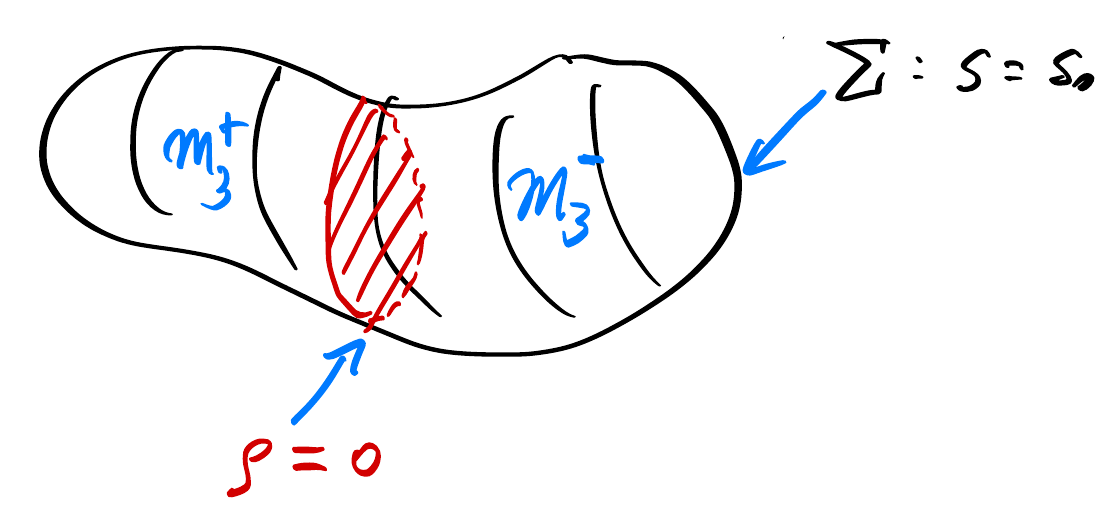}
    \caption{Sketch of a generic closed isentropic surface $\Sigma$ (defined by $s=s_0$), which must intersect the surfaces of zero charge density ($\rho=0$). The interior of the surface $\Sigma$ is thus split in two disjoint regions: $m_3^+$ where $\rho>0$, and $m_3^-$ where $\rho<0$.}
    \label{fig:isentropic}
\end{figure}\\
\textbf{Scenario 2.} A closed zero-charge density ($\rho=0$) surface $\widetilde{\Sigma}^0(t)$. Such a surface would exist generically, for a class of initial conditions. Because of the continuity equation, such a surface is advected by the velocity field. Without loss of generality, assume $\rho \geq 0$ in the interior of this surface. In this case, the conservation of mass gives
$$\widetilde{M} := \int_{\mathrm{int}(\widetilde{\Sigma}^0(t))}\rho \mathrm{d}^3x = \oint_{\widetilde{\Sigma}^0(t)} s \mbf{\omega}\cdot \hat{n} \mathrm{d}^2x\,,$$
again a boundary term.\\
}

{\subsubsection{Helicity conservation equation}\label{hce}
Helicity (or ``fluid helicity'') was originally introduced by \cite{moffatt1969degree} as the pseudo-scalar 
$H := \mathbf{V} \cdot \mbf{\omega}$, where $\mathbf{V}$ is the fluid velocity field and $\mbf{\omega} = \nabla \times \bt{V}$ is the  vorticity field. There, the conservation of helicity was demonstrated for barotropic flows (namely, when the pressure and mass density are related by $p=P(\rho_{\mathrm{mass}})$) in the dissipationless case. There is a good amount of literature about helicity conservation or non-conservation for inviscid compressible flows, and generalised invariants that follow from it. See for example \citep{bekenstein1987helicity,anco2020hierarchies}. However, it is instructive to obtain the fluid helicity and its conservation law starting from our proto-helicity, defined in equation \eqref{eq:j_heli_def} (or \eqref{eq:j_heli_def_2}), which satisfies the conservation law \eqref{eq:j_heli_div_ext_J}:
$$\partial_\mu \heli^\mu = 2 J^4\,, \qquad \heli^\mu = (A_a B^a, - A_0\,{\bf B} - {\bf A}\times {\bf E},0)\,.
$$
In terms of fluid variables, and choosing the gauge $\Phi \equiv 0$, this becomes
$$\partial_\mu \left.\heli^\mu\right|_{\Phi\equiv0} = 2 T \rho\,, \qquad \heli^\mu|_{\Phi\equiv0} = \frac{1}{\ellfive^2}\left(H, H \mathbf{V} + \left(h -\frac{1}{2}|\mathbf{V}|^2\right) \mbf{\omega}+ T \mathbf{V}\times \nabla s, 0\right)\,.
$$
\textbf{Results that apply to any dissipationless compressible $(3+1)$-dimensional fluid, irrespective of the constraint \eqref{eq:constraint}.} Notice that replacing $\rho$ in terms of the constraint \eqref{eq:constraint} and defining $H^\mu := \ellfive^2 \heli^\mu|_{\Phi\equiv0}$ gives a result that is independent of $\ellfive^2$:
\begin{equation}
    \label{eq:helicity}
\partial_\mu H^\mu = 2\, T \mbf{\omega}\cdot\nabla s\,, \qquad H^\mu = \left(H, H \mathbf{V} + \left(h -\frac{1}{2}|\mathbf{V}|^2\right) \mbf{\omega} + T \mathbf{V}\times \nabla s, 0\right)\,,
\end{equation}
which is in fact a known result \citep{bekenstein1987helicity}, valid for any dissipationless compressible $(3+1)$-dimensional fluid, irrespective of the constraint \eqref{eq:constraint}. The formulation \eqref{eq:helicity} highlights the role of the Rossby-Ertel invariant on the production of fluid helicity. However, as we will see below, helicity is conserved for barotropic fluids and is not necessarily conserved for baroclinic fluids. Because a fluid is barotropic if and only if $T\nabla s$ is irrotational, we consider a Helmholtz decomposition of $T\nabla s$ into an irrotational part (with zero curl) and a solenoidal part (with zero divergence), by which the `apparent' helicity production term $2\, T \mbf{\omega}\cdot\nabla s$ of equation \eqref{eq:helicity} splits into two terms, via the identity
$T (\nabla\times \mathbf{V})\cdot\nabla s \equiv \nabla \cdot (T \mathbf{V} \times \nabla s) + \mathbf{V} \cdot (\nabla T \times \nabla s)$. Equipped with this, the explicit form of equation \eqref{eq:helicity} becomes:
\begin{equation}
\label{eq:helicity2}
\partial_0 H + \nabla \cdot \left(  \bt{V} H  \right) = \nabla \cdot \left(  \left(\frac{1}{2}|\mathbf{V}|^2-h\right) \mbf{\omega} + T \mathbf{V}\times \nabla s  \right)+2\mathbf{V} \cdot (\nabla T \times \nabla s)\,,
\end{equation}
with the usual interpretation, via the Leibniz-Reynolds theorem \eqref{eq:Leibniz}, that the divergence term in the RHS of \eqref{eq:helicity2} provides a boundary contribution towards the time evolution of the integral 
\begin{equation}
\label{eq:helicity_integral}
    I(t):= \int_{m_3(t)}H\mathrm{d}^3x = \int_{m_3(t)}\bt{V}\cdot\mbf{\omega}\mathrm{d}^3x
\end{equation}
of fluid helicity over a volumetric region $m_3(t)$ that is advected by the fluid velocity field. We will call the term $2\mathbf{V} \cdot (\nabla T \times \nabla s)$ the `helicity production term'. Notice that this term is directly related to the `vorticity production term' $\nabla T \times \nabla s$ (a.k.a. baroclinity). \\ 
\textbf{Barotropic fluids.} In \citep{moffatt1969degree}, mainly two scenarios are considered: (i) barotropic fluids, namely where the mass density is a given function of the pressure: $\rho_{\mathrm{mass}} = R(p)$, and (ii) incompressible flows, where $\rho_{\mathrm{mass}} = \rho_0$ is constant and uniform. For each of these scenarios, the temperature $T$ becomes a function of the specific entropy $s$: $T=q'(s)$ for some function $q$, and thus the helicity production term $2\mathbf{V} \cdot (\nabla T \times \nabla s)$ becomes equal to zero throughout the spacetime. Next, under an appropriate choice of boundary conditions the total contribution from the boundary terms to the time derivative of $I(t)$ can be made equal to zero. More specifically, for barotropic fluids equation \eqref{eq:helicity2} simplifies to 
$$\partial_0 H + \nabla \cdot \left(  \bt{V} H  \right)  = \nabla \cdot \left(  \left(\frac{1}{2}|\mathbf{V}|^2 - h + q\right) \mbf{\omega}\right)\,,$$
and by choosing the volumetric region $m_3(t)$ so that the vorticity be parallel to $\partial m_3(t)$, as done in \citep{moffatt1969degree} (a condition that is preserved under the fluid flow evolution, since by virtue of the barotropic condition, the curl of $\frac{1}{\rho_{\mathrm{mass}}}\nabla p$ is equal to zero, and thus the vortex lines are stuck to the fluid as material lines), one gets via the Leibniz-Reynolds theorem:
\begin{equation}
\label{eq:deriv_helicity}
    \frac{\mathrm{d} I}{\mathrm{d} t} = \frac{\mathrm{d}}{\mathrm{d} t}\left(\int_{m_3(t)}H\mathrm{d}^3x\right) = \oint_{\partial m_3(t)} \left(\frac{1}{2}|\mathbf{V}|^2-h + q\right) \mbf{\omega}\cdot \mathbf{n} \mathrm{d}^2x = 0\,.
\end{equation}
\textbf{Barotropic fluids: Beyond Moffatt's boundary conditions.} It is important to stress that the above result relies on the boundary condition $\mbf{\omega}\cdot \mathbf{n}|_{\partial m_3(t)} \equiv 0$. Such a boundary condition has topological implications: for example, if $m_3(t)$ is compact, then vortex lines must wrap around inside $\partial m_3(t)$ and cannot pierce it. Thus, an obvious question is whether more general boundary conditions can be considered that generalise helicity conservation. It turns out that the proto-helicity can be used in this regard, by exploiting the gauge freedom. Let us introduce the gauge-dependent helicity as follows:
\begin{equation}
\label{eq:helicity_gauge}
    H_\Phi := \ellfive^2 \heli^0  = (\bt{V}-\ellfive \nabla \Phi)\cdot \mbf{\omega} = H  - \ellfive  \nabla \cdot(\Phi \mbf{\omega})\,,
\end{equation}
so this gauge-dependent helicity is equal to the classical fluid helicity plus a total divergence. Integrating over a general region $m_3(t)$ that is advected by the fluid without imposing any boundary condition, define
\begin{equation}
\label{eq:helicity_integral_gauge}
    I_\Phi(t) := \int_{m_3(t)}H_\Phi\mathrm{d}^3x = I(t) - \ellfive \oint_{\partial m_3(t)} \Phi \mbf{\omega}\cdot \mathbf{n}\mathrm{d}^2x\,.
\end{equation}
Taking time derivative we  get two contributions. First, $\frac{\mathrm{d} I}{\mathrm{d} t} = \oint_{\partial m_3(t)} \left(\frac{1}{2}|\mathbf{V}|^2-h + q\right) \mbf{\omega}\cdot \mathbf{n} \mathrm{d}^2x$, which is not necessarily zero. Second,  
\begin{equation}
\label{eq:deriv_gauge}
    \frac{\mathrm{d} }{\mathrm{d} t}\left(\ellfive \oint_{\partial m_3(t)} \Phi \mbf{\omega}\cdot \mathbf{n}\mathrm{d}^2x\right) = 
\ellfive \oint_{\partial m_3(t)} (\partial_0 + \mathcal{L}_{\bt{V}}) \Phi \mbf{\omega}\cdot \mathbf{n}\mathrm{d}^2x + \ellfive \oint_{\partial m_3(t)} \Phi (\partial_0 + \mathcal{L}_{\bt{V}}) \mbf{\omega} \cdot \mathbf{n}\mathrm{d}^2x\,,
\end{equation}
where $\mathcal{L}_{\bt{V}} \Phi = \bt{V}\cdot \nabla \Phi$, cf. equation \eqref{eq:Lie_A_4}, and  $\mathcal{L}_{\bt{V}}\mbf{\omega} = (\bt{V}\cdot \nabla)\mbf{\omega} + (\nabla\cdot \bt{V})\mbf{\omega} - (\mbf{\omega}\cdot\nabla) \bt{V}$, cf. equation \eqref{eq:Lie_B}. Now, from the curl of the momentum equation \eqref{V2} we obtain $(\partial_0 + \mathcal{L}_{\bt{V}}) \mbf{\omega} = \nabla T \times \nabla s = 0$ for barotropic fluids, so combining the two time derivatives we get
$$\frac{\mathrm{d} I_\Phi}{\mathrm{d} t} = \oint_{\partial m_3(t)} \left(\frac{1}{2}|\mathbf{V}|^2-h + q - \ellfive (\partial_0 + \mathcal{L}_{\bt{V}}) \Phi\right) \mbf{\omega}\cdot \mathbf{n} \mathrm{d}^2x\,,$$
which generalises the fluid helicity invariant to the case of arbitrary boundary conditions, provided the gauge field $\Phi$ satisfies a first-order linear inhomogeneous ODE along pathlines:
$$\partial_0 \Phi + \bt{V}\cdot\nabla \Phi = \frac{1}{\ellfive}\left(\frac{1}{2}|\mathbf{V}|^2-h + q\right) \quad \text{on} \quad {\mathcal{M}}_4 \quad  \Longrightarrow \quad \frac{\mathrm{d} I_\Phi}{\mathrm{d} t} = 0\,.$$
In summary, this new invariant can be interpreted as a generalization, for barotropic fluids, of the boundary conditions on the vorticity field, giving a formula for the `old' fluid helicity:
\begin{equation}
\label{eq:new_helicity}
    \int_{m_3(t)}\bt{V}\cdot \mbf{\omega}\mathrm{d}^3x = I_\Phi +  \oint_{\partial m_3(t)} (\ellfive\Phi) \mbf{\omega}\cdot \mathbf{n}\mathrm{d}^2x\,,
\end{equation}
where $I_\Phi=$constant and $\ellfive\Phi$ satisfies $\partial_0 (\ellfive\Phi) + \bt{V}\cdot \nabla (\ellfive\Phi) = \frac{1}{2}|\mathbf{V}|^2-h + q$.
We note that such a gauge choice is equivalent to the condition $A_\mu V^\mu = A_0 + \bt{A}\cdot \bt{V} + A_4 V^4 =  \frac{1}{\ellfive} \left(s q'(s) -q(s)\right)$.\\
\textbf{Baroclinic fluids.} When the barotropic condition is dropped and the mass density depends on pressure and temperature, the analysis made in \citep{moffatt1969degree} does not apply: we expect the helicity production term  $2\mathbf{V} \cdot (\nabla T \times \nabla s)$ to be nonzero somewhere, and not to be expressible as a total divergence. However, it is instructive to study this case in light of the gauge-dependent helicity defined above for barotropic fluids. Going back to equations \eqref{eq:helicity2}--\eqref{eq:helicity_integral}, which define the evolution of the fluid helicity $H$, we readily obtain
$$\frac{\mathrm{d} I}{\mathrm{d} t}  = \oint_{\partial m_3(t)} \left(\frac{1}{2}|\mathbf{V}|^2-h\right) \mbf{\omega}\cdot \mathbf{n}\mathrm{d}^2x + \oint_{\partial m_3(t)} T (\mathbf{V}\times \nabla s)  \cdot \mathbf{n}\mathrm{d}^2x + \int_{m_3(t)}2\mathbf{V} \cdot (\nabla T \times \nabla s)\mathrm{d}^3x\,,$$
and we see two types of boundary contributions: one proportional to $\mbf{\omega}\cdot \mathbf{n}$ and another proportional to $(\mathbf{V}\times \nabla s)  \cdot \mathbf{n}$. While these are proportional to each other in the barotropic case, they are not proportional in the baroclinic case. Now, defining the gauge-dependent helicity as in the barotropic case, namely via equations \eqref{eq:helicity_gauge}--\eqref{eq:helicity_integral_gauge}, we get
$$\frac{\mathrm{d} I_{\Phi}}{\mathrm{d} t}  =\frac{\mathrm{d} I}{\mathrm{d} t} - \frac{\mathrm{d} }{\mathrm{d} t}\left(\ellfive \oint_{\partial m_3(t)} \Phi \mbf{\omega}\cdot \mathbf{n}\mathrm{d}^2x\right)\,,$$
and using equations \eqref{eq:deriv_gauge} and $(\partial_0 + \mathcal{L}_{\bt{V}}) \mbf{\omega} = \nabla T \times \nabla s$ (not necessarily zero now) gives
$$\frac{\mathrm{d} }{\mathrm{d} t}\left(\ellfive \oint_{\partial m_3(t)} \Phi \mbf{\omega}\cdot \mathbf{n}\mathrm{d}^2x\right) = 
\ellfive \oint_{\partial m_3(t)} (\partial_0 + \mathcal{L}_{\bt{V}}) \Phi \mbf{\omega}\cdot \mathbf{n}\mathrm{d}^2x + \ellfive \oint_{\partial m_3(t)} \Phi (\nabla T \times \nabla s) \cdot \mathbf{n}\mathrm{d}^2x\,,$$
so combining these we obtain
\begin{eqnarray}
\nonumber
    \frac{\mathrm{d} I_{\Phi}}{\mathrm{d} t} &=& \oint_{\partial m_3(t)} \left(\frac{1}{2}|\mathbf{V}|^2-h - \ellfive (\partial_0 + \mathcal{L}_{\bt{V}}) \Phi\right) \mbf{\omega}\cdot \mathbf{n}\mathrm{d}^2x + \oint_{\partial m_3(t)} ((T \mathbf{V} -\ellfive \Phi \nabla T)\times \nabla s)  \cdot \mathbf{n}\mathrm{d}^2x\\ 
\label{eq:deriv_baroclinic_helicity}
    &+& \int_{m_3(t)}2\mathbf{V} \cdot (\nabla T \times \nabla s)\mathrm{d}^3x\,.
\end{eqnarray}
At this point, to simplify the analysis we consider a scenario where the advected region $m_3(t)$ is such that $s|_{\partial m_3(t)} = s_0$ (a constant)%, cf. figure \ref{fig:isentropic}
. Such a boundary condition is preserved in time because the specific entropy $s$ is a conserved scalar. In this case, we have $\nabla s \propto \mathbf{n}$ on  $\partial m_3(t)$, so the second boundary integral in \eqref{eq:deriv_baroclinic_helicity} vanishes. Thus, to eliminate the remaining boundary term we set
$$ \partial_0\Phi + \bt{V}\cdot\nabla \Phi =\frac{1}{\ellfive} \left(\frac{1}{2}|\mathbf{V}|^2-h \right)  \quad \text{on} \quad {\mathcal{M}}_4 \quad  \Longrightarrow \quad \frac{\mathrm{d} I_\Phi}{\mathrm{d} t} = \int_{m_3(t)}2\mathbf{V} \cdot (\nabla T \times \nabla s)\mathrm{d}^3x\,.$$
\textbf{A result that depends on the constraint \eqref{eq:constraint}.} We now consider the production of helicity in the above scenario, but using the constraint \eqref{eq:constraint} as an extra equation. Using again the identity  $T (\nabla\times \mathbf{V})\cdot\nabla s \equiv \nabla \cdot (T \mathbf{V} \times \nabla s) + \mathbf{V} \cdot (\nabla T \times \nabla s)$ and the condition $s|_{\partial m_3(t)} = s_0$ (a constant), we get
$$ \frac{\mathrm{d} I_{\Phi}}{\mathrm{d} t} = \int_{m_3(t)}2\mathbf{V} \cdot (\nabla T \times \nabla s)\mathrm{d}^3x =  \int_{m_3(t)}2 T \mbf{\omega}\cdot\nabla s\mathrm{d}^3x = \int_{m_3(t)}2\ellfive^2 T \rho\mathrm{d}^3x\,,$$
and recalling from the analysis in Scenario 1 of subsection \ref{sec:mass_cons}, cf. figure \ref{fig:isentropic}, that the interior of an isentropic surface necessarily contains regions with positive $\rho$ and regions with negative $\rho$, namely we can separate $m_3(t) = m_3^+(t) \cup m_3^-(t)$ with $\pm\rho > 0$ on $m_3^\pm(t)$, we get
$$ \frac{\mathrm{d} I_{\Phi}}{\mathrm{d} t} =  \int_{m_3^+(t)}2\ellfive^2 T \rho_{\mathrm{mass}}\mathrm{d}^3x - \int_{m_3^-(t)}2\ellfive^2 T \rho_{\mathrm{mass}}\mathrm{d}^3x\,,$$
clearly indicating the two competing contributions to this gauge-dependent helicity.}

{
\subsection{Examples based on an ideal fluid: Scenarios of rotating fluids} 
For an ideal fluid, $e = c_v T$ and $p = R \rho T$, with $c_v$ and $R$ positive constants related to specific properties of the fluid. Notice that these formulae are extended directly to the case $\rho<0$. Solving equations \eqref{-p}--\eqref{-S/N}, we get 
\begin{equation}
\label{eq:ideal_gas}
f = T[c_v[ 1 - \ln (T/T_0)] + R \ln (|\rho|/\rho_0)]\,, \qquad s = c_v \ln (T/T_0) - R \ln (|\rho|/\rho_0)\,.
\end{equation}
Because vorticity plays an important role in our formulation, we will consider scenarios of rotating fluids, and will simplify the study by looking for  axisymmetric solutions. To this end, we will use cylindrical coordinates $({r},z)$ (${r} \geq 0$ is the radial coordinate) such that the fluid's angular velocity $\mbf{\Omega}:=\Omega \hat{z}$ be constant and parallel to the $z$-axis.  Neglecting self-gravity, the gravity field is imposed simply by inserting an external potential energy function $\Psi(\mbf{x})$ into the action principle, which amounts to the following redefinition of the enthalpy in the momentum equation \eqref{V2}: $h \to h+\Psi$. We leave the form of the potential energy $\Psi(\mbf{x})$ free for the moment. \\
Thus, the initial velocity field is $\bt{V} = \mbf{\Omega}\times \mbf{x}$, so the initial vorticity field is $\mbf{\omega} = \nabla \times \bt{V} = 2 \mbf{\Omega} = \Omega \hat{z}$.
\subsubsection{Steady solution for an axisymmetric rotating barotropic fluid (rotating cylinder)}
We look for a time-independent, axisymmetric solution for all fields. Let us first look at the continuity equation for the charge density \eqref{rho_cons}:
$\nabla\cdot(\bt{V} \rho) = 0$, or simply $\bt{V} \cdot \nabla \rho = 0$, namely ${\Omega} {r} \partial_\varphi \rho = 0$, where $\varphi$ is the azimuthal angle. This equation is satisfied identically due to our assumption of axisymmetry. Next, the constraint \eqref{eq:constraint} reads $\ellfive^2 \rho = 2 \Omega \partial_z s$, which determines the charge density $\rho$ in terms of the specific entropy $s$. Next, equation \eqref{SH/N}, corresponding to conservation of entropy, is automatically satisfied. Finally, we look at equations \eqref{V2}, corresponding to momentum conservation. We have $\partial_0 \bt{V} = 0$ and $(\bt{V}\cdot \nabla)\bt{V} = {r} \Omega^2 \partial_\varphi (\hat{\varphi}) = -{r} \Omega^2 \hat{{r}}$, so the balance reads 
\begin{equation}
\label{eq:balance_rotating_cylinder}
    -{r} \Omega^2 \hat{{r}}  = -\nabla\left(\frac{1}{2}\Omega^2 {r}^2\right) = -\nabla(h+\Psi) + T \nabla s\,.
\end{equation}
This is the only equation that remains to be considered. Taking the curl of it we obtain a necessary condition: $T = T(s)$, which defines a barotropic fluid. Next, the ideal gas law \eqref{eq:ideal_gas} can be solved for $|\rho|$:
\begin{equation}
\label{eq:rho_ideal_gas}
    |\rho| =\rho_0 \exp\left(-\frac{s}{R} + \frac{c_v}{R} \ln (T(s)/T_0)\right) = \rho_0 \left(\frac{T(s)}{T_0}\right)^\frac{c_v}{R}\exp\left(-\frac{s}{R}\right)\,.
\end{equation}
But recall that we had $\ellfive^2 \rho = 2 \Omega \partial_z s$ from the constraint. Squaring this we get finally
$$4 \Omega^2 (\partial_z s)^2 =\ellfive^4\rho^2 = \ellfive^4\rho_0^2 \left(\frac{T(s)}{T_0}\right)^\frac{2c_v}{R}\exp\left(-\frac{2s}{R}\right)  \,,$$
which has an interpretation of a $1$-dimensional particle with zero energy and with potential energy function 
$$U(s) \propto - \left(\frac{T(s)}{T_0}\right)^\frac{2c_v}{R}\exp\left(-\frac{2s}{R}\right)\,,$$
which in principle may have turning points, at the zeros of $T(s)$ (which are also the zeros of $\rho$ because of the ideal gas law \eqref{eq:rho_ideal_gas}). Thus, if the barotropic fluid is such that $T(0) = T(s_{\max}) = 0$ and $T(s) > 0$ for $0< s < s_{\max}$ throughout the fluid (with $T(s) = 0$ for $s<0$ or $s>s_{\max}$), then we obtain an equation of the form
\begin{equation}
\label{eq:1D-particle_s}
    \partial_{zz} s = - U'(s)
\end{equation}
with two turning points, with solution $s(z)$ periodic in $z$ with half-period $L$ representing the vertical length of the cylinder. Fixing the initial condition  at the left turning point to have coordinate $z=-L/2$, recalling that $\rho \propto \partial_z s$ we would then have $\rho > 0$ for $-L/2<z<L/2$. For $|z|\geq L/2$ we could set $\rho=T=0$. The ``barotropic relation'' would look like $p \propto |\rho|^{1+R/c_v} \mathrm{e}^{s/c_v}$, very similar to the usual barotropic relation for adiabatic fluids.\\
Finally, going back to equation \eqref{eq:balance_rotating_cylinder} it remains to solve for the potential $\Psi$. From $h=f+Ts+p/\rho$ and using the ideal gas relations \eqref{eq:ideal_gas}, we get $h = (c_v+R) T$ and introducing the anti-derivative $T(s) := q'(s)$, equation \eqref{eq:balance_rotating_cylinder} reduces to the scalar equation
\begin{equation}
\label{eq:balance_Psi}
    - \frac{1}{2}\Omega^2 {r}^2 = -(c_v+R) q'(s) + q(s) - \Psi\,,
\end{equation}
which determines the potential energy function $\Psi$ in terms of the spatial coordinates $({r},z)$ once $s(z)$ is known from the solution of equation \eqref{eq:1D-particle_s}.\\
In conclusion, it is possible to find a steady solution of our equations in a rotating cylinder scenario. Notice that the shape of $\Psi$ would be feasible experimentally as it would have a main linear component as a function of $z$ due to the function $q(s)$ being monotonically increasing with $s$, and due to $s$ being monotonically increasing with $z$.\\
Finally, the helicity of this configuration is identically zero, because $H = \bt{V}\cdot\mbf{\omega} = 0$ throughout spacetime. Notice that the conservation of helicity would follow the steps leading to equation \eqref{eq:deriv_helicity}, but with $h \to h + \Psi$, so we would get
$$\frac{\mathrm{d}}{\mathrm{d} t}\left(\int_{m_3(t)}H\mathrm{d}^3x\right) = \oint_{\partial m_3(t)} \left(\frac{1}{2}|\mathbf{V}|^2-h - \Psi + q\right) \mbf{\omega}\cdot \mathbf{n} \mathrm{d}^2x,$$
which is equal to zero --regardless of any boundary conditions on $\mbf{\omega}$-- thanks to the defining equation for $\Psi$, equation \eqref{eq:balance_Psi}. 
\subsubsection{Vorticity production in a baroclinic rotating planet}
We saw in the last example that the force balance equation \eqref{eq:balance_rotating_cylinder} is the challenging one. Taking the curl of its RHS  gives the vorticity production (a.k.a.~\emph{baroclinity}) at $t=0$:
$$\left.\partial_0 \mbf{\omega}\right|_{t=0} = \left.\nabla T \times \nabla s\right|_{t=0}\,.$$
We see immediately that for barotropic fluids this production vanishes and thus a steady solution exists as per the last example. However, in reality the atmosphere is not barotropic and thus it is worth considering a baroclinic scenario, which is by definition unsteady. Letting our fields describe the atmosphere of a rotating spherical planet and letting the origin of coordinates $({r},z)$ be at the centre of the planet, using the same notation as in the last example, we look for axisymmetric solutions. In the remainder of this subsection all quantities refer to the initial state $t=0$. We see from equation \eqref{eq:ideal_gas} that $\nabla T \times \nabla s = \frac{R T}{c_v} (\nabla \ln \rho) \times \nabla s$, so using the constraint $\ellfive^2 \rho = 2 \Omega \partial_z s$ we get
$$\nabla T \times \nabla s = \frac{R T}{c_v \partial_z s} (\nabla \partial_z s) \times \nabla s = \frac{R T}{c_v \partial_z s} \left(\partial_{zz} s \partial_{r} s - \partial_{z} s \partial_{z{r}} s\right) \hat\varphi\,,$$
which shows that baroclinity implies $\partial_{{r}} s \neq 0$ somewhere in the fluid. In fact, in such baroclinic regions we can write
$$\nabla T \times \nabla s =  \frac{R T}{c_v} (\partial_{r} s) \,\partial_z \ln \left(\frac{\partial_{z} s}{\partial_{r} s}\right) \hat\varphi\,.$$
Now, letting $\Theta({r},z)$ be the angle that $\nabla s({r},z)$ makes with the positive $z$-axis, we have $\frac{\partial_{z} s}{\partial_{r} s} = \cot \Theta$, so 
$$\nabla T \times \nabla s =  -\frac{R T}{c_v} (\partial_z s) \,(\partial_z \tan\Theta)\,  \hat\varphi\,,$$
so the baroclinity condition is simply $\partial_z \tan\Theta \neq 0$, namely $\nabla s$ is not parallel transported along the $z$-axis. For example, this baroclinity condition would be realised if $s$ was a function of the distance to the centre of the planet. Letting $s = F({r}^2+z^2)$ for some differentiable function $F$ we get $\tan \Theta = {\partial_{{r}} s}/{\partial_z s} = {r}/z$ (i.e., $\Theta = \theta$, the polar angle), and thus
$$\nabla T \times \nabla s = -\frac{R T}{c_v} (\partial_z s) \,\partial_z \left(\frac{{r}}{z}\right)\,  \hat\varphi = 2\frac{R T}{c_v} \frac{{r}}{z} F'({r}^2+z^2) \,  \hat\varphi\,,$$
but  the ideal gas law \eqref{eq:rho_ideal_gas} gives $T=T_0 \left(\frac{|\rho|}{\rho_0}\right)^{\frac{R}{c_v}}\exp\left(\frac{s}{c_v}\right)$ and the constraint gives $\rho = 2 \Omega \partial_z s / \ellfive^2 = 4 \Omega z F'({r}^2+z^2) / \ellfive^2$, so
\begin{equation}
\label{eq:baroclinity_planet}
    \nabla T \times \nabla s = 2  \mathrm{sign}(z) {{|z|}^{\frac{R}{c_v}-1}} \frac{R T_0}{c_v} \left(\frac{4 \Omega}{\rho_0 \ellfive^2}\right)^{\frac{R}{c_v}}{{r}} F'({r}^2+z^2) \left({|F'({r}^2+z^2)|}\right)^{\frac{R}{c_v}}\mathrm{e}^{\frac{s}{c_v}}\,  \hat\varphi\,.
\end{equation}
We can now study this vorticity production term (baroclinity) physically. Let us assume a scenario where entropy grows with height (i.e. $F'>0$), corresponding physically to a dry atmosphere in the troposphere. Thus, the baroclinity points in the $\hat{\varphi}$ direction in the northern hemisphere (i.e. for $z>0$). Next, at a given distance to the centre of the planet, the baroclinity changes direction when going from $z$ to $-z$, namely, when going from the northern hemisphere to the southern hemisphere at the same reference latitude. Thus, this mechanism will produce a Ferrel cell-like pattern in the atmosphere, namely the type of atmospheric circulation cell typical of  mid-latitudes, which is unaffected by solar heating or polar cooling \citep{ferrel1856essay}. Figure \ref{fig:planet} shows an example of the initial atmospheric mass density profile $\rho_{\mathrm{mass}} = 2 \Omega \partial_z s /\ellfive^2$ on a planet in the region $3 < x^2+y^2+z^2 < 4$, with initial entropy profile $s \propto 1-\exp(-3(x^2+y^2+z^2 - 3))$, which is monotonically increasing as a function of the distance to the centre of the planet. Of course, this example simply shows that initially the flow is going to depart from the steady bulk rotation and will begin forming atmospheric circulation cell-like patterns. The axisymmetric character of the solution may subsequently be lost due to instability. The subsequent evolution of this configuration is worth studying numerically in future work.\\
Finally, we notice that the baroclinity \eqref{eq:baroclinity_planet} has a singularity at $z=0$, namely at the equator, because $\frac{R}{c_v}<1$ (physically the singularity is due to the fact that the mass density goes to zero at $z=0$). However, this singularity is integrable in the sense that the production of vorticity flux (a.k.a.~production of circulation) is finite in a given hemisphere. This is easy to see, as the integral $\int_0^{z_0} {|z|}^{\frac{R}{c_v}-1} \mathrm{d}z$ converges due to the fact that $\frac{R}{c_v} > 0$. In a similar fashion, the initial production of helicity, being proportional to $\bt{V}\cdot(\nabla T \times \nabla s)$ (with $\bt{V} \parallel \nabla T \times \nabla s$) will be nonzero and will have a singularity at $z=0$, which again is integrable so the integral form of helicity production \eqref{eq:deriv_baroclinic_helicity} produces a finite change.
}

\begin{figure}
    \centering
    \includegraphics[width=0.35\linewidth]{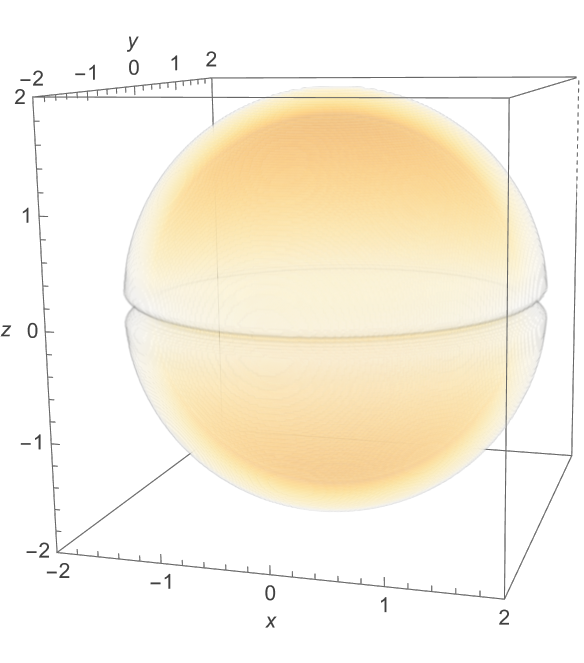}\hspace{5mm} \includegraphics[width=0.07\linewidth]{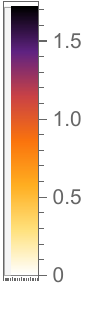}
    \caption{{For the rotating planet example, 3D density plot of the initial mass density $\rho_{\mathrm{mass}} = \left|2 \Omega \partial_z s /\ellfive^2\right| \propto |z| \exp(-3(x^2+y^2+z^2 - 3))$, for an initial entropy of the form $s \propto 1-\exp(-3(x^2+y^2+z^2 - 3))$,   in the region $3 < x^2+y^2+z^2 < 4$. The baroclinity points in the $\hat{\varphi}$ ($-\hat{\varphi}$) direction in the northern (southern) hemisphere, as in a Ferrel cell.} \label{fig:planet}}
\end{figure}

\section{Conclusions} 
\label{sec:conclusions}

We have analysed the classical Abelian Chern-Simons theory in $4+1$ spacetime dimensions as fluid model. The action principle provides advection equations and an advecting velocity field that is obtained directly as the null vector of the field strength (curvature) tensor. In the free theory, prototypical versions of helicity and entropy densities arise. In the interacting theory  with a minimal coupling to an external current source, the evolution equations preserve the advection formulation: the external current determines both the advecting velocity and a conserved scalar density, which is the prototype of the Rossby-Ertel's potential vorticity charge density. Finally, in the self-interacting theory we identify the external current with the Chern-Simons connection, via the introduction of non-relativistic kinetic and internal energies in the action principle, and using a `dressing' gauge field to ensure the gauge invariance of the theory. The resulting equations, when reduced to $3+1$ spacetime dimensions, become equivalent to the evolution equations for a classical dissipation-free compressible $(3+1)$-dimensional fluid, endowed with thermodynamics, but with a constraint on the initial conditions: the potential vorticity charge must be of uniform magnitude, which is determined by the length of the fifth dimension.

{In this interpretation of the Chern-Simons theory as a fluid system, the prototypical versions of helicity, entropy and mass density provide new insights regarding the conservation of fluid helicity, mass and entropy in dissipationless fluids. The constraint provided by the Chern-Simons formulation on the Rossby-Ertel's potential vorticity is interpreted as a particular choice of initial condition: $\ellfive^2 \rho_{\mathrm{mass}} = |\mbf{\omega} \cdot \nabla s|$, which forces the fluid to have significant vorticity $\mbf{\omega}$ and entropy gradients $\nabla s$ in regions of high mass density $\rho_{\mathrm{mass}}$. Thus, the type of flows that this theory describes are extremely active, making the theory relevant in scenarios of significant rotation and convection, as in planets and stars. Specifically for atmosphere, the Chern-Simons model forces the Rossby-Ertel's potential vorticity to take piecewise constant values $\pm \ellfive^2$ in regions separated by zero-mass-density advected surfaces. This make sense because, as we have shown (cf. figure \ref{fig:isentropic}), isentropic surfaces must intersect these zero-mass-density surfaces, and we can look at the evidence from atmospheric data: there, on isentropic surfaces potential vorticity tends to attain high-magnitude values within advected regions in jet streams or during events such as extreme cyclogenesis, of common occurrence above midlatitude seas in winter \citep{haynes1990conservation, mcintyre2003potential}.
Conversely, this Chern-Simons model, due to its constraint, cannot describe flow regimes where matter is not rotating, or where matter does not have a spatial modulation in its thermodynamical variables. Hence, it does not describe simple cases such as uniform motion, isentropic flow, or potential flow. But we provide non-trivial steady solutions of the theory for a rotating ideal gas, and we also show, for a rotating planet's dry atmosphere, that baroclinity (i.e. vorticity production) is necessarily present, in a Ferrel cell-like pattern.} 

In terms of results that apply to any dissipationless compressible  $(3+1)$-dimensional fluid (namely, without a constraint on the potential vorticity charge), a new fluid helicity invariant is proposed which generalises the classical fluid helicity \citep{moffatt1969degree} to the case where the vorticity is not necessarily tangent to the boundary of the advected region $m_3(t) \subset \mathbb{R}^3$. This invariant, based on the gauge dependence of the proto-helicity, has important applications. For example, for 3D Beltrami flows the velocity and the vorticity are given time-independent functions defined on a $3$-dimensional space, such that $\mbf{\omega} = \lambda \bt{V}$ on the spatial manifold ${\mathcal{M}}_3$ (usually taken as $\mathbb{R}^3$ or $\mathbb{T}^3$), where $\lambda$ is a nonzero numerical constant. Hence, Moffatt's boundary condition $\mbf{\omega}\cdot \bt{n}|_{\partial m_3} = 0$ implies $\bt{V}\cdot \bt{n}|_{\partial m_3} = 0$, namely the velocity field is parallel to the $2$-dimensional manifold ${\partial m_3}$, and hence this manifold is invariant under the flow. Thus, ${\partial m_3}$ must be an invariant torus of the flow, which exists but is very difficult to find: for example, it is almost surely knotted \citep{Enciso_Peralta-Salas_Romaniega_2023}. Now, in practical applications, given the initial region $m_3(0)$ and its boundary ${\partial m_3}(0)$, Moffatt's boundary condition $\mbf{\omega}\cdot \bt{n}|_{\partial m_3} = 0$ is not satisfied initially, and thus it is never satisfied. Moreover, the evolution of the boundary $\partial m_3$, being a ``material'' surface, is almost surely chaotic \citep{Enciso_Peralta-Salas_Romaniega_2023}. Our new invariant is thus the natural generalisation of the fluid helicity invariant, leading to formula \eqref{eq:new_helicity} for the `old' fluid helicity.

{Our full $(4+1)$-dimensional Chern-Simons model has an extra dimension, with local coordinate $x^4$. We can use this to model dissipation in the $(3+1)$ spacetime, in a conservative way because the current must be conserved in the full $(4+1)$ spacetime. This will be explored in a subsequent paper, particularly in terms of modelling phase changes and other emergent phenomena.}

{We based our self-interacting Chern-Simons theory on a non-relativistic kinetic energy and internal energy prescription, which gave us the equations of motion for a non-relativistic dissipationless compressible fluid. The theory and the equations of motion can be made relativistic trivially, simply by replacing the energy terms with their relativistic counterparts. We will explore this in future work, particularly in connection with stellar environments, accretion disks near black holes, and the interaction between a fluid and an electromagnetic field, as well as in terms of non-Abelian extensions.}

{A word about gauge invariance in fluid dynamics. Our formulation is in accordance with the Clebsch formulation \citep{salmon1988hamiltonian, zakharov1997hamiltonian, morrison1998hamiltonian}, in that in a generic fluid, the velocity field is usually related to an external reference frame, possibly ``the  container''  that is supposed to be absolute and inertial, but this is actually an arbitrary choice. The formulation of the theory should in no way depend on the existence of that frame and the physical phenomena described by the fluid equations should be fully covariant under changes of reference frame; they should be equally valid if referred to the container or with respect to the frame moving with a fish. In fact, the interactions between parcels of the fluid should have the same meaning for someone at rest with respect to the container, co-moving with the fluid, or otherwise. This freedom means that the description should be invariant if the velocity field $\bt{V}$ is replaced by $\bt{V}+ \bt{u}(\bt{x})$, where $\bt{u}$ can be identified with $\nabla \Omega(\bt{x})$ for some arbitrary scalar gauge function $\Omega$. This can be viewed as the origin of the gauge freedom $\bt{A} \to \bt{A}' =\bt{A} + \mathrm{d}\Omega$ in fluid dynamics.}

\section{Acknowledgments}
M.~D.~B.~acknowledges support by Science Foundation Ireland (SFI) under Grant Number 12/IP/1491 and by INFN, {and is grateful for the hospitality of the DISAT of the Torino Polytechnic.} J.~Z. is grateful for the hospitality of CASL at University College Dublin and the DISAT of the Torino Polytechnic, where most of this work was carried out. This work has been partially funded through FONDECYT Grants 1220862, 1230112, 1230492 and 1241835, and by INFN.

\section{Author Contributions}
 All authors contributed equally to this work.

\appendix
\section{Notations and conventions}

\subsection{Space-time metric, $p$-forms and Hodge dual}\label{notations metric}
We adopt a mostly plus signature of space-time:
$\eta_{\mu\nu} := \mathrm{diag}(-,+,+,+,+)$.
Furthermore,
$$\epsilon^{01234}=+1=-\epsilon_{01234}\,.$$
We use, for $p$-forms, the convention
$$\Omega^{(p)} :=\frac 1{p!}\Omega_{\mu_1\cdots \mu_p}\,\mathrm{d}x^{\mu_1}\wedge \cdots \wedge\mathrm{d}x^{\mu_p}\,,$$
so that $\mathrm{F}\equiv \frac 12 \mathrm{F}_{\mu\nu}\mathrm{d}x^\mu\wedge \mathrm{d}x^\nu$.
Furthermore,
we define electric and magnetic fields by
$${E_a := \mathrm{F}_{a0}\,,\qquad B_a := \frac{1}{2}\epsilon_{abc} \mathrm{F}_{bc} \qquad (\text{thus\,\,\,} \mathrm{F}_{ab}= \epsilon_{abc}B_c)\,.}$$
We define the Hodge dual in $1+4$ dimensions:
$$ {}^* \Omega^{(p)} := \frac 1{p!(5-p)!}\Omega_{\mu_1\cdots \mu_p} \mathrm{d}x^{\nu_1}\wedge \cdots \wedge \mathrm{d}x^{\nu_{5-p}}\epsilon_{\nu_1\cdots \nu_{5-p}}{}^{ \mu_1\cdots\mu_p}\,,$$
so that
$$\Omega^{(p)}\wedge {}^*\Xi^{(p)}=\frac{(-1)^{(5-p)p+1}}{p!}\,\Omega_{\mu_1\cdots \mu_p}\,\Xi^{\mu_1\cdots \mu_p}\,\mathrm{d}^5 x\,.$$
In particular, the action contains the term:
$$A\wedge{}^*J=-A_\mu J^\mu\,\mathrm{d}^5x\,.$$

{\subsection{Lie derivatives}
\label{sec:Lie_der}
Let $\bt{V}$ be a $3$-dimensional vector field defined on a $3$-dimensional spatial manifold $\mathcal{M}_3$. Then the following local formulae for Lie derivatives (along $\bt{V}$) of tensors of various ranks defined on $\mathcal{M}_3$ hold:
\begin{itemize}
    \item For a $0$-form $A_4$ (i.e. a scalar field covariant under coordinate transformations):
\begin{equation}
    \label{eq:Lie_A_4}
    \mathcal{L}_{\bt{V}} A_4 = \bt{V} \cdot \nabla A_4\,.
\end{equation}
    \item For a $0$-form density $\rho$:
\begin{equation}
    \label{eq:Lie_rho}
    \mathcal{L}_{\bt{V}} \rho = \nabla \cdot (\bt{V} \rho)\,.
\end{equation}
    \item For a  $1$-form $\bt{A} := A_a \mathrm{d}x^a$:
\begin{equation}
    \label{eq:Lie_A}
    \mathcal{L}_{\bt{V}} \bt{A} = (\bt{V} \cdot \nabla) \bt{A} + A_a \nabla V^a\,.
\end{equation}
    \item For a vector field $\bt{U} := U^a \frac{\partial}{\partial x^a}$:
\begin{equation}
    \label{eq:Lie_U}
    \mathcal{L}_{\bt{V}} \bt{U} = (\bt{V} \cdot \nabla) \bt{U} - (\bt{U} \cdot \nabla) \bt{V}\,.
\end{equation}
    \item For a vector field density $\bt{B} := B^a \frac{\partial}{\partial x^a}$ (such as the magnetic field or the vorticity field):
\begin{equation}
    \label{eq:Lie_B}
    \mathcal{L}_{\bt{V}} \bt{B} = \frac{\partial}{\partial x^a} (V^a \bt{B}) - (\bt{B} \cdot \nabla) \bt{V}\,.
\end{equation}
\end{itemize}
}

\subsection{Eigenvectors problem}\label{eigen}
Let us discuss here the null eigenvectors $V^\mu$ of the $\mathrm{F}_{\mu\nu}$ tensor in the free case{, satisfying the condition
\begin{align}
    V^\mu \mathrm{F}_{\mu\nu}=0\,.
\end{align}}
Using the $SO(4)$ symmetry we can always reduce $\mathrm{F}_{ij}$ ($i,j=1,\ldots,4$) to  the following normal form
\begin{align}
\mathrm{F}_{ij}=    \begin{pmatrix}
0& \mathrm{F}_{12}& 0&0\cr
-\mathrm{F}_{12}& 0&0&0\cr
0&0&0&\mathrm{F}_{34}
\cr
0&0&-\mathrm{F}_{34}& 0\cr\end{pmatrix}\,.
\end{align}
The above matrix is still invariant under the residual $SO(2)\times SO(2)$ symmetry, each factor acting on the couples of indices $1,2$ and $3,4$ separately.

We can use this symmetry to further fix, say:
$$\mathrm{F}_{02}=\mathrm{F}_{04}=0\,.$$
We finally end up we the following normal form for $\mathrm{F}_{\mu\nu}$, to hold in a given point of space-time:
\begin{align}\label{normal}
\mathrm{F}_{\mu\nu}=   \begin{pmatrix}
0& \mathrm{F}_{01}& 0 &\mathrm{F}_{03}&0 \cr
-\mathrm{F}_{01}&0& \mathrm{F}_{12}& 0&0\cr
0&-\mathrm{F}_{12}& 0&0&0\cr
-\mathrm{F}_{03}&0&0&0&\mathrm{F}_{34}
\cr
0&0&0&-\mathrm{F}_{34}& 0\cr\end{pmatrix}\,.
\end{align}
In the free case, the condition
$$\mathrm{F}\wedge \mathrm{F}=0$$
implies:
\begin{align}
\mathrm{F}_{12}\mathrm{F}_{34}=0\,, \quad \mathrm{F}_{01}\mathrm{F}_{34}=0
\quad \text{and} \quad
\mathrm{F}_{12}\mathrm{F}_{03}=0\end{align}
Let us choose $\mathrm{F}_{34}=0$.

Solving the null eigenvector equation $V^\mu \mathrm{F}_{\mu\nu}=0$, we have 2 possibilities:
\begin{enumerate}
\item $\mathrm{F}_{03}=0$, in which case:
$$V^\mu=\left(V^0,0,\frac{\mathrm{F}_{01}}{\mathrm{F}_{12}}V^0,V^3,V^4
\right)$$
\item $\mathrm{F}_{12}=0$, in which case:
\begin{enumerate}
\item if $\mathrm{F}_{03}\neq 0$:
$$V^\mu=\left(0,V^1,V^2,-\frac{\mathrm{F}_{01}}{\mathrm{F}_{03}}V^1,V^4
\right)$$
\item $\mathrm{F}_{03}= 0$ and $\mathrm{F}_{01}\neq 0$:
$$V^\mu=\left(0,0,V^2,V^3,V^4
\right)$$
\end{enumerate}
\end{enumerate}
All the cases depend, modulo an overall factor, only on 2 independent parameters. 
Only in the case (i) we can have the time component $V^0\neq 0$, in which case $V^\mu$ can be associated with a conserved current or with a velocity vector. Any two non-space-like vectors in this class can always be normalised so as to differ by a 2-parameter space-like vector  (e.g. the cases with $V^4=0$ and $V^4\neq 0$ are mutually independent).
In the case (ii), the vector $V^\mu$ is necessarily space-like.
Note that, in case we choose a two-time  signature, the analysis changes significantly.\par
In the interacting case $J^\mu\neq 0$, if $J^0\neq 0$, referring to the normal form \eqref{normal} of $\mathrm{F}_{\mu\nu}$, we have $\mathrm{F}_{12} \mathrm{F}_{34}\neq 0$ and, recalling eq. \eqref{A}, the matrix  has a single 0-eigenvector of the form
\begin{equation}J^\mu=(\mathrm{F}_{12} \mathrm{F}_{34}, 0, \mathrm{F}_{01} \mathrm{F}_{34}, 0, \mathrm{F}_{03} \mathrm{F}_{12})\,.\end{equation}
As opposed to the free case, now we can uniquely characterise the velocity of the fluid as $J^\mu/J^0$.

\section{Covariant derivation of the proto-helicity advection}
\label{app:heli_covariant}

In the notation of section \ref{sec:pHelicityCurrent}, using equation \eqref{conservation_law_PE} we can consider a region ${\mathcal{M}}_4$ of the four-dimensional spacetime spanned by the spatial region $\Omega$ during the evolution of the fluid from time $t$ to $t+\mathrm{d}t$ and write 
\begin{equation}
    0=\int_{{\mathcal{M}}_4}\,\partial_{\hat\mu}{\heli}^{\hat\mu}= \int_{\partial {\mathcal{M}}_4}{\heli}_{\hat\mu} {n}^{\hat{\mu}}\,\mathrm{d}^3\sigma\,,\label{consH}
\end{equation}
where ${n}^{\hat{\mu}}$ is the outward-directed unit $4$-vector normal to the boundary $\partial {\mathcal{M}}_4$ of ${\mathcal{M}}_4$. This boundary, in turn, consists of $\Omega(t+\mathrm{d}t)$, $\Omega(t)$ and of a 3-dimensional  region $\widetilde{\partial M}$ connecting the boundaries $\partial \Omega(t)$ and $\partial\Omega(t+\mathrm{d}t)$, consisting of infinitesimal flows, with velocity $V^{\hat{\mu}}$,  originating at the points of $\partial\Omega(t)$.

The unit normal vector and integration measure on this latter region can be written as:\footnote{We choose the ``mostly plus'' convention for the metric, $\epsilon^{01234}=+1=-\epsilon_{01234}$ and $\epsilon^{\hat{\mu}\hat\nu\hat\rho\hat\sigma}:=\epsilon^{\hat{\mu}\hat\nu\hat\rho\hat\sigma 4}$.}
$$\left.n^{\hat{\mu}}\,\mathrm{d}^3\sigma\right\vert_{\widetilde{\partial M}}=\frac{1}{2}\left.\epsilon^{\hat \mu\hat\nu\hat\rho\hat\sigma}\,V_{\hat{\nu}} \mathrm{d}x_{\hat{\rho}}\wedge  \mathrm{d}x_{\hat{\sigma}}\wedge \mathrm{d}t\right\vert_{\partial\Omega(t)}\,.$$
One can indeed easily verify that $V_{\hat\mu}\,n^{\hat{\mu}}=0$ on $\widetilde{\partial M}$.
We can then expand the boundary integral on the right-hand side of \eqref{consH} as follows:
\begin{align}
    0&=\mathcal{H}(t+\mathrm{d}t)-\mathcal{H}(t)+\int_{\partial\Omega(t)} {\heli}_{\hat\mu}\mathrm{d}\boldsymbol{\sigma}^{\hat{\mu}}\,\mathrm{d}t=\frac{\mathrm{d}}{\mathrm{d}t}\mathcal{H}\,\mathrm{d}t+\int_{\partial\Omega(t)}{\heli}^0\,({\bf V}'-{\bf V})\cdot \hat{n}\,\mathrm{d}S\,\mathrm{d}t\,,\label{DHDt}
\end{align}
where $\hat{n}$ is the outward-directed normal to $\partial\Omega(t)$ and $\mathrm{d}S$ the surface element on it. Using the Leibniz-Reynolds transport theorem, the time derivative of $\mathcal{H}$ can, in turn, be written as follows:
\begin{equation}
    \frac{\mathrm{d}}{\mathrm{d}t}\mathcal{H}=\int_{\Omega(t)}(\partial_0+\mathcal{L}_{\bf V})({\bf A}\wedge \mathrm{d}{\bf A})=\int_{\Omega(t)}\left(\partial_t {\heli}^0\right)\,\mathrm{d}^3x+\int_{\partial\Omega(t)}{\heli}^0\,{\bf V}\cdot \hat{n}\,\mathrm{d}S\,.
\end{equation}
Note that equation \eqref{DHDt} 
is analogous to the condition of vanishing of the ``material derivative'' $D\mathcal{H}/Dt$ of a conserved quantity $\mathcal{H}$, which is being computed on a \emph{control volume} $\Omega(t)$ whose boundary points move at velocity ${\bf V}(t)$, while the quantity  $\mathcal{H}$ itself is transported at velocity ${\bf V}'(t)\neq {\bf V}(t)$.
Equation \eqref{DHDt} can then be written as follows:
\begin{equation}
    \frac{\mathrm{d}}{\mathrm{d}t}\mathcal{H}=\int_{\Omega(t)}\left(\partial_t {\heli}^0\right)\,\mathrm{d}^3x+\int_{\partial\Omega(t)}{\heli}^0\,{\bf V}\cdot \hat{n}\,\mathrm{d}S=-\int_{\partial\Omega(t)}{\heli}^0\,({\bf V}'-{\bf V})\cdot \hat{n}\,\mathrm{d}S\,.
\end{equation}
The proto-helicity $\mathcal{H}$ is conserved if:
$$\int_{\partial\Omega(t)}{\heli}^0\,({\bf V}'-{\bf V})\cdot \hat{n}\,\mathrm{d}S=0\,.$$
One solution is ${\heli}^0{\bf V}'={\heli}^0{\bf V}$, which is generally too restrictive. However, we can choose $\partial\Omega(t)$ to be a surface of uniform entropy: $\left.A_4\right\vert_{\partial\Omega(t)}\equiv {\rm const.}$ In this case, $\hat{n}\propto\nabla A_4$ and 
$${\heli}^0({\bf V}'-{\bf V})\cdot \nabla A_4=0\,,$$
so that the surface integral vanishes. To prove the above relation, let us write the components $\nu=4$ of the equations \eqref{JF} and \eqref{VF}:
\begin{align}
    {\heli}^a\,\mathrm{F}_{a4}+{\heli}^0\,\mathrm{F}_{04}&=0\,,\nonumber\\
    {V}^a\,\mathrm{F}_{a4}+\mathrm{F}_{04}&=0\,.
\end{align}
{Multiplying the latter} equation by ${\heli}^0$ and subtracting them side by side, we find
$${\heli}^0(V^{\prime a}-V^a)\,\mathrm{F}_{a4}={\heli}^0(V^{\prime a}-V^a)\,\partial_a A_4=0\,.$$
In the presence of an external source $J^\mu$ equation \eqref{conservation_law_PE} no longer holds.  We have 
$$\partial_{\hat\mu}{h}^{\hat\mu}=\frac{1}{4}\,\epsilon^{\hat\mu\hat\nu\hat\rho\hat\sigma}\mathrm{F}_{\hat\mu\hat\nu}\mathrm{F}_{\hat\rho\hat\sigma}={2}\,J^4\,.$$
Integrating both sides over ${\mathcal{M}}_4$, we find that the substantive derivative of $\mathcal{H}$ now reads:
\begin{equation}
\frac{D\mathcal{H}}{Dt}=\frac{\mathrm{d}\mathcal{H}}{\mathrm{d}t}+\int_{\partial\Omega(t)}{h}^0\,({\bf V}'-{\bf V})\cdot \hat{n}\,\mathrm{d}S={2}\,\int_{\Omega(t)}J^4\,\mathrm{d}^3x\,.
\end{equation}
Moreover, equation \eqref{JF} changes into:
\begin{equation}
 {{h}}^{\hat{\mu}}\,\mathrm{F}_{\hat{\mu}\nu}={\left(A_{\hat{\nu}}\,J^4,\,-A_{\hat{\mu}}J^{\hat{\mu}}\right)}\,,
\end{equation}
so that
\begin{equation}
 {{h}}^0 ({\bf V}'-{\bf V})\cdot \nabla A_4={-\,A_{\hat{\mu}}J^{\hat{\mu}}}\,.
\end{equation}
{\section{Advection of surfaces with constant $s$ and $\rho$}\label{lastappendix}
In this Appendix, we characterise the time evolution, within the fluid, of surfaces with constant $s$ ($\Sigma_{s_0}(t)$) and $\rho$ ($\Sigma_{\rho_0}(t)$):
$$\left. s\right\vert_{\Sigma_{s_0}(t)}=s_0={\rm const.}\,\,,\,\,\left. \rho\right\vert_{\Sigma_{\rho_0}(t)}=\rho_0={\rm const\,.}
$$
Let us consider $\Sigma_{s_0}(t)$ first and describe a generic point on it, defined by intrinsic coordinates $\xi=(\xi^{\alpha})$ on the surface, by the position vector ${\bf x}(t,\xi)$.
The advection velocity of the surface ${\bf V}_{\rm adv}$ is defined as:
$$\frac{\mathrm{d} }{\mathrm{d}t}{\bf x}(t,\xi)={\bf V}_{\rm adv}(t,\xi)\,.$$
The equation defining $\Sigma_{s_0}(t)$ is:
$$s(t,\,{\bf x}(t,\xi))=s_0={\rm const.}$$
Deriving both sides with respect to time we find:
$$\left.(\partial_0 s+{\bf V}_{\rm adv}\cdot \nabla s)\right\vert_{\Sigma_{s_0}(t)}=0\,.$$ 
From the above equation, using eq. \eqref{SH/N}, we find
\begin{equation}\left.({\bf V}_{\rm adv}-{\bf V})\cdot \nabla s\right\vert_{\Sigma_{s_0}}=0\,,\label{sconst}\end{equation}
which implies that ${\bf V}_{\rm adv}$ and ${\bf V}$ have the same normal component to the surface. We can say that $\Sigma_{s_0}$ is advected by the velocity ${\bf V}$ of the fluid.\par
Consider now $\Sigma_{\rho_0}(t)$ defined by the equation:
$$\rho(t,\,{\bf x}(t,\xi))=\rho_0={\rm const.}$$
the definition of ${\bf x}(t,\xi)$ and of ${\bf V}_{\rm adv}$ being the same as for $\Sigma_{s_0}(t)$. Deriving both sides with respect to time, we find:
$$\left.(\partial_0 \rho+{\bf V}_{\rm adv}\cdot \nabla \rho)\right\vert_{\Sigma_{\rho_0}(t)}=0\,.$$ 
Using now the continuity equation \eqref{rho_cons}, we find
\begin{equation}\left.(-\rho_0\,\nabla\cdot {\bf V}+
({\bf V}_{\rm adv}-{\bf V})\cdot\nabla \rho)\right\vert_{\Sigma_{\rho_0}(t)}=0\,.\label{rhoconst0}\end{equation}
We would find the same property of $\Sigma_{s_0}(t)$ only if the first term on the left-hand side vanishes. We conclude that, for a compressible fluid, only the surface at $\rho=\rho_0=0$ is advected at the velocity of the fluid.\par
Consider now the integral
$$M(t)=\int_{m_3(t)}\rho\,\mathrm{d}^3{ x}\,,$$ and let us evaluate its time derivative.
Using the Leibniz-Reynolds theorem and the continuity equation for $\rho$, we find:
\begin{equation}
    \frac{\mathrm{d}M}{\mathrm{d}t}=\oint_{\partial m_3} \rho\,({\bf V}_{\rm adv}-{\bf V})\cdot {\bf n}\,\mathrm{d}^2x\,.
\end{equation}
In light of eqs. \eqref{sconst}, \eqref{rhoconst0}, the right-hand side vanishes if one of the following conditions are satisfied: $\partial m_3$ is a surface at constant $s$; $\partial m_3$ is a surface at $\rho=0$; $\partial m_3$ is the union of a surface at constant $s$ and a surface at $\rho=0$. The last case corresponds to the volumes $m^\pm_3$ considered in section \ref{hce}. In general, from the above discussion, we can conclude that $M(t)$ is conserved if it is computed on a volume $m_3(t)$ whose boundary is advected by the velocity ${\bf V}$ of the fluid.
}

%\bibliographystyle{jfm}
%\bibliography{jfm}
%Use of the above commands will create a bibliography using the .bib file. Shown below is a bibliography built from individual items.

\bibliographystyle{jfm}
\bibliography{jfm}

\end{document}